\documentclass[prd,aps,12pt]{revtex4}
\usepackage{graphicx}
\usepackage{amsmath}
\usepackage{amssymb}
\usepackage{epsfig,psfrag}

\newcommand{\Ub}{\bar{U}}
\newcommand{\Db}{\bar{D}}
\newcommand{\zb}{\bar{z}}

\newcommand{\omegab}{\bar{\omega}}

\newcommand{\wb}{\bar{w}_{1}}
\newcommand{\w}{{w}_{1}}
\newcommand{\Jm}{\mathcal{J}}
\newcommand{\Dm}{\mathcal{D}}
\newcommand{\dellr}{\overleftrightarrow{\partial}}
\newcommand{\Tr}{\mathrm{Tr}}
\renewcommand{\theequation}{\arabic{section}.\arabic{equation}} 

\begin{document}
\title{Effective field theory for Sp(N) antiferromagnets \\
 and its phase structure}
\author{Keisuke Kataoka, Shinya Hattori, and Ikuo Ichinose}
\affiliation{Department of Applied Physics, Nagoya Institute of Technology,
Nagoya, Japan}

\begin{abstract}
In this paper, we study quantum Sp(N) antiferromagnetic (AF) Heisenberg
models in two dimensions (2D) by using the Schwinger-boson representation
and the path-integral methods.
An effective field theory, which is an extension of CP$^{N-1}$ model in (2+1)D,  
is derived and its phase structure is studied by the $1/N$-expansion.
We introduce a spatial anisotropy in the exchange couplings 
and show that the effective coupling
constant in the CP$^{N-1}$ model is an increasing function of the 
anisotropy.
For the SU(N) AF Heisenberg model, which is a specific case of the
Sp(N) model, we found that phase transition from the ordered ``N\'eel state" to 
paramagnetic phase takes place as the anisotropy is increased.
In the vicinity of the SU(N) symmetric point, this phase structure is 
retained.
However as a parameter that controls explicit breaking of the SU(N)
symmetry is increased, a new phase, which is
similar to the spiral-spin
phase with a nematic order in frustrated SU(2) spin systems, appears.
It is shown that at that phase transition point, a local SU(2) gauge
symmetry with composite SU(2) gauge field appears in the low-energy sector.
It is another example of symmetry-enhancement phenomenon at low energies.
We also introduce a lattice gauge-theoretical model,
which is a counterpart of the effective field theory, and study its
phase structure by means of the Monte-Carlo simulations. 
\end{abstract}
\maketitle
\section{Introduction}

Study on quantum antiferromagnets has been one of the most active
area in the condensed matter physics.
In particular in the last decade,
much attention has been paid to exotic phase and exotic phase transition for
which Landau's classic paradigm cannot be applicable\cite{glw}.
It is expected that investigation of such exotic states is useful to understand
anomalous properties of under doped high-$T_c$ materials\cite{lnw}.
Furthermore recent development in technologies of ultracold atoms
and optical lattice trap elevates purely academic quantum spin models
to realistic ones, and these cold-atom systems are sometimes
regarded as a final simulator for strongly-correlated electron systems.
Quantum SU(N) antiferromagnets are one of these examples.
Theoretically these models can be studied by using the Schwinger-boson
methods and the $1/N$-expansion\cite{rs}.
In a seminal paper\cite{cold}, it was shown that spin-$(N-1)/2$ ($N$ is an even
integer) cold atom systems in an optical lattice with one atom per 
quantum well can be regarded as quantum Sp(N) antiferromagnets. 
There are two parameters $J_1,\; J_2$ in the Hamiltonian 
of the Sp(N) magnets and when $J_1=J_2$, the symmetry is enhanced to 
SU(N)$\supset$Sp(N) and a SU(N) quantum antiferromagnet is realized.

In the present paper, we shall study Sp(N) quantum antiferromagnets 
by using the slave-boson (Schwinger boson) representation and the path-integral
methods.
We first derive an effective field theory for the Sp(N) Heisenberg
model.
This field theory is an extension of the CP$^{N-1}$ model for the SU(N) 
antiferromagnetic (AF) Heisenberg model.
Then we investigate its phase structure by using the 
$1/N$-expansion.
We also study numerically its lattice-gauge-model counterpart 
by means of the Monte-Carlo (MC) simulations.

This paper is organized as follows.
In Sect.2, we shall derive the effective field theory for the Sp(N)
AF Heisenberg model by using CP$^{N-1}$ representation
of (pseudo-)spin degrees of freedom.
In Sect.3, we study the field-theory model by the $1/N$-expansion.
We focus on quantum phase transition in the spatial two-dimensional 
(2D) system.
We first show that in the SU(N) AF magnets with anisotropic
exchange couplings on a square lattice, a phase transition from
the AF N\'eel state to the paramagnetic state takes place as the anisotropy
is increased.
These two states persist in the Sp(N) system in the vicinity of the SU(N) 
symmetric point $J_1=J_2$.
As $J_2/J_1$ is decreased to some critical value, 
a phase transition to a new phase with a
composite vector-field condensation takes place and a new kind of spin
order appears.
In Sect.4, we study a lattice version of the obtained field theory and
show the results of the numerical study for the Sp(4) case.
Obtained phase diagram is qualitatively in agreement with that obtained 
by the $1/N$-expansion, but we also find some discrepancy 
between the $1/N$-expansion and the numerical study, e.g. order
of the phase transition, etc.
Section 5 is devoted for conclusion.
In the Appendix, some details of the derivation of the effective field theory
are given.

\section{Model and effective field theory}
\setcounter{equation}{0}
\subsection{Sp(N) Heisenberg model}

We consider anisotropic Sp(N) Heisenberg model on a square lattice, i.e.,
exchange couplings in the $x$ and $y$-directions are different.
Quantum Hamiltonian of the system is given as
follows in the most general form, 
\begin{equation}
{\cal H}=\sum_{\langle i,j \rangle }
\Big\{ J^{i,j}_1 \sum_{a,b}\Gamma^{ab}_i \Gamma^{ab}_j
-J^{i,j}_2 \sum_a\Gamma^a_i \Gamma^a_j \Big\},
\label{spnha}
\end{equation}
where $i, j$ denote lattice sites and 
$\Gamma^{ab} \in {\bf sp} (\mathrm{N})$ (the Lie algebra
of Sp(N)) have ${N(N+1) \over 2}$ components.
On the other hand, 
$\Gamma^{a} \in {\bf su}(\mathrm{N}) / {\bf sp} (\mathrm{N})$,
and there are ${(N+1)(N-2) \over 2}$ components.
Hereafter in most of cases, we set the exchange coupling $J_{1}^{i,j}$ 
and $J_{2}^{i,j}$ as follows,
$$
J_{n}^{i,i+\hat{x}} = J_{n,x}, \; J_{n}^{i,i+\hat{y}} = J_{n,y}, \;
(n=1,2), \; \mbox{otherwise $0$},
$$
where $\hat{x}\; (\hat{y})$ is the unit vector of the $x$($y$)-direction.
For the $N=4$ case\cite{qi}, the above generators are explicitly given as 
\begin{equation}
\Gamma^{a} = \sigma^{a} \otimes \mu^{z}, a = 1,2,3, \;\;
\Gamma^{4} = 1 \otimes \mu^{x}, \;\;
\Gamma^{5} = 1 \otimes \mu^{y}, 
\label{eqb}
\end{equation}
and 
\begin{equation}
\Gamma^{ab} = \frac{1}{2i}[\Gamma^{a},\Gamma^{b}], 
\label{eqc}
\end{equation}
where $\sigma^{a},\mu^{a} (a = 1,2,3)$ are two sets of the $2 \times 2$ Pauli matrices.
The Sp(4) case corresponds to the spin-3/2 system, and $\Gamma^a$ involves 
even powers of the SU(2) spin matrices, whereas $\Gamma^{ab}$ involves
odd powers of the spin matrices.
Then the long-range order of $\Gamma^a$ indicates a spin-nematic order. 

It is useful to introduce
the matrix $\Jm$, which has the following properties and 
is a generalization of the time-reversal matrix $i\sigma^2$ in the SU(2) spin case,
\begin{equation}
\Jm^{t} = -\Jm, \ \Jm^{2} = -1, \ 
\Jm\Gamma^{ab}\Jm = (\Gamma^{ab})^{t}, \ 
\Jm\Gamma^{a}\Jm = -(\Gamma^{a})^{t}. 
\label{eqd}
\end{equation}
For the Sp(4) case with Eqs.(\ref{eqb}), $\Jm = i\sigma^{y} \otimes \mu^{x}$.

We introduce the Schwinger boson operator 
$\hat{b}_\alpha\; (\alpha=1, \cdots, N)$
and represent the ``spin operators" in the Hamiltonian (\ref{spnha}) 
in terms of them, 
$
\hat{\Gamma}^{ab}_{i}=\hat{b}_{i,\alpha}^{\dagger}\Gamma^{ab}_{\alpha\beta}
\hat{b}_{i,\beta}$, \;
$\hat{\Gamma}^{a}_{i}=\hat{b}_{i,\alpha}^{\dagger}\Gamma^{a}_{\alpha\beta}
\hat{b}_{i,\beta}.
$
By using the identities,
\begin{align}
\sum_{a,b}\Gamma^{ab}_{\alpha\beta}\Gamma^{ab}_{\gamma\sigma} 
&= 2\delta_{\alpha\sigma}\delta_{\beta\gamma} - 
2\Jm_{\alpha\gamma}\Jm_{\beta\sigma} \label{eq:5}, \\
\sum_a\Gamma^{a}_{\alpha\beta}\Gamma^{a}_{\gamma\sigma} 
&= 2\delta_{\alpha\sigma}\delta_{\beta\gamma} + 
2\Jm_{\alpha\gamma}\Jm_{\beta\sigma}
- \delta_{\alpha\beta}\delta_{\gamma\sigma}, \label{eq:6}
\end{align}
we obtain
\begin{equation}
\mathcal{H}=\sum_{\langle i,j\rangle}\{2(J_{1}^{i,j}-J_{2}^{i,j})
\hat{K}_{ij}^{\dagger}\hat{K}_{ij}
-2(J_{1}^{i,j}+J_{2}^{i,j})\hat{Q}_{ij}^{\dagger}
\hat{Q}_{ij} \},
\label{spnhb}
\end{equation}
where 
$\hat{Q}_{ij} = \Jm_{\alpha\beta}\hat{b}_{i,\alpha}\hat{b}_{j,\beta}$
and 
$\hat{K}_{ij} = \hat{b}^{\dagger}_{\alpha,i}\hat{b}_{\alpha,j}$.
The operator $\Jm_{\alpha\beta}\hat{b}_{j,\beta}$ is the
conjugate spinor of $\hat{b}_{j,\beta}$ and then
$\hat{Q}_{ij}$ represents pairing of spins on the lattice sites
$i, j$, whereas $\hat{K}_{ij}$ corresponds to the Schwinger boson
(spinon) hopping.
Subsidiary condition 
\begin{equation}
\sum_{\alpha=1}^N \hat{b}^\dagger_{i,\alpha} \hat{b}_{i,\alpha}
|\mbox{phys}\rangle=|\mbox{phys}\rangle,
\label{subcon}
\end{equation}
must be imposed as the physical-state condition.
We redefine the exchange couplings as 
$J_{i,j} \equiv 2(J_{1}^{i,j}+J_{2}^{i,j})$,
$J'_{i,j} \equiv 2(J_{1}^{i,j}-J_{2}^{i,j})$,
then 
\begin{equation}
{\cal H}=\sum_{\langle i,j\rangle}\{J'_{i,j}\hat{K}_{ij}^{\dagger}\hat{K}_{ij}
-J_{i,j}\hat{Q}_{ij}^{\dagger}\hat{Q}_{ij} \}.
\label{spnhc}
\end{equation}
From Eq.(\ref{spnhc}), it is obvious that when $J'_{i,j}=0$
($J_{i,j}=0$), the model have the global SU(N) symmetry.
In the following section, we shall derive the effective field theory 
for the system (\ref{spnhc}) by using the path-integral methods.

\subsection{Effective field theory}

In this subsection, we shall derive the low-energy effective field theory of
the Hamiltonian (\ref{spnhc}).
To this end, we use the coherent path-integral methods.
For the Schwinger boson, we use the CP$^{N-1}$ boson that satisfies
$\bar{z}_i\cdot z_i=\sum_\alpha \bar{z}_{i,\alpha}z_{i,\alpha}=1$
corresponding to the condition (\ref{subcon}).
Then the partition function is given as
\begin{align}
Z &=\int\mathcal{D}\zb\mathcal{D}z \delta(\zb \cdot z -1)
\exp \left[\int^{\beta}_{0}d\tau A(\tau) \right] \notag \\
A(\tau) &= -\sum_{i,\alpha}
\zb_{i,\alpha}\dot{z}_{i,\alpha} -\mathcal{H}(z,\zb), 
\label{za}
\end{align}
where $\mathcal{H}(z,\zb)$ is obtained from Eq.(\ref{spnhc}) by replacing 
$\hat{b}_{i,\alpha}\rightarrow {z}_{i,\alpha}$ and
$\hat{b}^\dagger_{i,\alpha}\rightarrow \bar{z}_{i,\alpha}$.
\begin{figure}[th]
\begin{center}
\includegraphics[width=7cm]{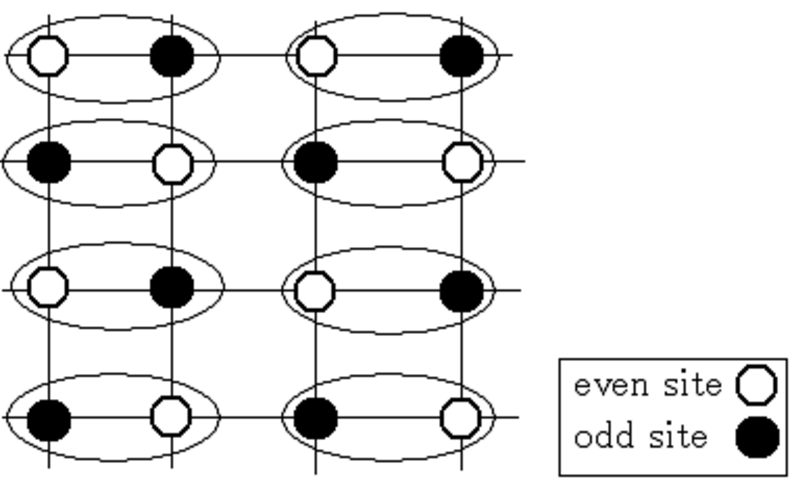}
\end{center}
\caption{CP$^1$ variables on odd sites are paired with their NN $z_e$.}
\label{aa}
\end{figure}
In order to obtain the effective field theory from Eq.(\ref{za}),
we integrate out the half of the CP$^{N-1}$ variables, e.g. those at
odd sites\cite{im}.
To this end, we introduce a complete-orthogonal set of vectors
in the CP$^{N-1}$ space, $\{ w_i; i=0,1,\cdots, N-1\}$.
Arbitrary CP$^{N-1}$ variable $z$ can be expanded as 
$z=\sum_ip_iw_i$, where $p_i$'s are complex numbers satisfying 
$\sum_i|p_i|^2=1$.
As we are interested in the case with the both exchange couplings
 $J_{i,j},\; J'_{i,j}\geq 0$,
dominant configurations in the path integral are given by 
$z_e \approx z$ at even sites and $z_o \approx \Jm \bar{z}$ at odd sites,
where $z$ is a smoothly varying CP$^{N-1}$ field.
We parameterize odd-site $z_o$ by referring to one of its nearest-neighbor(NN) 
even sites $z_e$ as (see Fig.\ref{aa}), 
\begin{equation}
z_o=\sum_i p_i w_i,
\label{zo}
\end{equation}
where $w_0=z_e,\; w_1=\Jm \bar{z}_e$ and we do not have to specify the
other $w_i$'s.
Then by the above remarks on the dominant configurations, 
we can assume $p_i (i\neq 1)\ll 1$ and 
\begin{align}
p_{1} 
&= U(1-\sum_{i\neq 1}|p_{i}|^{2})^{1/2} \notag \\
&\approx U - \frac{1}{2}U(\sum_{i\neq 1}|p_{i}|^{2}),  
\label{pa}
\end{align}
where $U$ is a U(1) variable that appears as a result of 
the local U(1) symmetry of the system, 
$z_{i,\alpha} \rightarrow e^{i\theta_i}z_{i,\alpha}$.

\begin{figure}[th]
\begin{center}
\includegraphics[width=7cm]{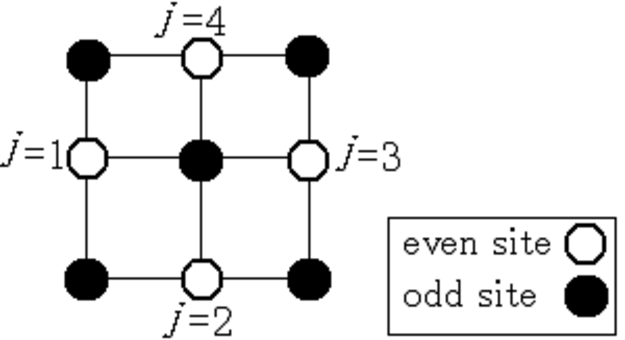}
\end{center}
\caption{$z_o$ and its NN $z_j$'s on even sites.}
\label{ab}
\end{figure}

From Eq.(\ref{pa}), 
\begin{equation}
z_{o} \approx \sum_{i \ne 1}p_{i}w_{i} + 
U\left( 1 - \dfrac{1}{2}\sum_{i \ne 1}|p_{i}|^{2}\right)w_{1}.
\label{zob}
\end{equation}
By substituting Eq.(\ref{zob}) into the action (\ref{za}),
we obtain
\begin{align}
A(\tau) 
=\sum_{\rm odd-site}(A_{0}
+\bar{\mathbf{p}}\cdot \mathbf{k}
+\bar{\mathbf{l}} \cdot \mathbf{p}
-\bar{\mathbf{p}} \mathbf{M} \mathbf{p} ) +O(\mathbf{p}^4),
\label{aaa}
\end{align}
where $\mathbf{p}=(p_0,p_2,p_3, \cdots, p_{N-1})^t$.
The matrix $\mathbf{M}$ is explicitly given as follows,
\begin{equation}
\mathbf{M} = \left( \begin{array}{cccc}
-4J_{o,j}-4J'_{o,j} & 0         & \cdots & 0 \\
0                  & -4J_{o,j} & \cdots & 0 \\
\cdot               & 0       & \cdot   & \cdot  \\
\cdot               & \cdot       & \cdot   & 0  \\
0                  & 0         & \cdots    & -4J_{o,j} 
\end{array}
\right)
\label{m}
\end{equation}
and the ``vectors" $\mathbf{k}$ and $\bar{\mathbf{l}}$ have rather
complicated form of composite of $z$, $\bar{z}$, $w$ and $\bar{w}$, 
which are explicitly shown in the Appendix.
In Eq.(\ref{m}), $J_{o,j}\; (J'_{o,j})$ denotes $J_{i,j} \; (J'_{i,j})$ for 
$i=o$ and the NN even $j$ (see Fig.\ref{ab}).
As the kernel $\mathbf{M}$ has negative diagonal elements, the 
Gaussian integral over $\mathbf{p}$ can be safely done,
\begin{align}
Z 
&=\int\mathcal{D}\zb\mathcal{D}z \delta(\zb \cdot z -1)
\exp \left[\int^{\beta}_{0}d\tau A(\tau) \right] \notag\\
&=\int\mathcal{D}\bar{\mathbf{p}}\mathcal{D}\mathbf{p} 
\exp \left[\int^{\beta}_{0}d\tau A(\tau) \right] \notag\\
&\approx \exp
\bigg[
\int_{0}^{\beta} d \tau A_{z}(\tau)
\bigg],
\label{zb}
\end{align}
\begin{equation}
A_{z}(\tau) = \sum_{\mathrm{odd-site}} 
\bar{\bf{l}}(\tau){\bf{M}}^{-1}(\tau){\bf{k}}(\tau) 
+ \sum_{\mathrm{odd-site}}A_{0}(\tau),
\label{ac}
\end{equation}
\begin{equation}
A_{0} = 
-\sum_{j}J_{o,j}(\bar{z}_{j}\mathcal{J}\bar{w}_{1})(z_{j}\mathcal{J}w_{1})
-\sum_{j}J'_{o,j}(\wb z_{j})(\zb_{j}\w),
\label{ax}
\end{equation}
where $\sum_j$ in Eq.(\ref{ax}) denotes the summation over even site $j$ 
around odd site $o$ (see Fig.\ref{ab}).

We can consider a continuum limit of the effective model (\ref{ac}),
as the remaining variables $z_j$'s can be regarded as a smoothly varying field
$z(r)$ ($r_0=\tau, r_1=x, r_2=y$).
Hereafter we explicitly set the exchange couplings between adjacent spins as follows,
\begin{equation}
J_{x}=J_{0}, \ J_{y}=\lambda J_{0}, \ J'_{x}=J'_{0}, \ 
J'_{y}=\lambda J'_{0}, 
\label{js}
\end{equation}
where $\lambda$ is the parameter for the anisotropy between $x$ and $y$
directions.
We take the limit $\beta={1 \over T} \rightarrow \infty$, introduce explicitly the
lattice spacing $a$ and rescale the coordinates as,
\begin{align}
\begin{cases}
\tau \rightarrow \sqrt{2}(1+\lambda)J_{0}a\tau =c\tau  \\
x \rightarrow \sqrt{\frac{1+\lambda}{2}}x \\
y \rightarrow \sqrt{\frac{1+\lambda}{2\lambda}}y, 
\label{rescaling}
\end{cases}
\end{align}
where $c$ is the ``speed of light" of the system.
Then we obtain action of the effective field theory for anisotropic Sp(N) 
AF magnets in two dimensions,
\begin{equation}
S= \frac{1}{2g}\int d^{3}r 
\Big(D_{\mu} \zb D_{\mu}z-\gamma(\zb \Jm D_{\mu}\zb)(z \Jm D_{\mu}z) 
+\sigma (|z|^{2} - 1)\Big),
\label{sa}
\end{equation}
where $D_\mu=\partial_\mu-\bar{z}\partial_\mu z$,
$\sigma$ is the Lagrange multiplier for the CP$^{N-1}$ constraint,
$\gamma$ is an increasing function of $J'_0/J_0$,
$\gamma=J'_0/J_0+O((J'_0/J_0)^2)$, and the coupling 
constant $g$ is given by
\begin{equation}
g = \frac{1+\lambda}{\sqrt{2\lambda}}a.
\label{g}
\end{equation}
From Eq.(\ref{g}), it is obvious that the effective coupling $g$ has
the minimum at the isotropic point $\lambda=1$.
As we see in the following section,
this means that the anisotropy tends to break the AF order of the
ground state.

As it is shown in the Appendix, there also exists a Berry phase in the action,
\begin{equation}
S_{B} =- \frac{1}{6}
\sqrt{\frac{2\lambda}{1+\lambda}}\Big(1-{J'_0 \over J_0}\Big)
\int d^{3}r
\epsilon_{\mu\nu\rho}D_\mu(\Db_{\nu}\zb D_{\rho}z). 
\label{sb}
\end{equation}
The Berry phase (\ref{sb})
with the fractional coefficient depending on 
the anisotropy $\lambda$ and $J'$-coupling does not suppress
effects of instanton in contrast to that with an integer coefficient,
i.e., $S_B$ does not give any substantial effects on the 
phase structure and critical behavior\cite{yoshioka}.
For the case of the SU(2) AF magnets on 2D lattice with anisotropic couplings,
this observation has been verified directly by the numerical study\cite{Janke}.

In the following sections, we shall study the field theory (\ref{sa})
by analytical and numerical methods.

\section{Phase structure: Analytical study}
\setcounter{equation}{0}
\subsection{$1/N$ expansion: Case of small $\gamma$}

The partition function of the effective field theory is given as,
\begin{align}
Z=\int \Dm z \Dm \zb \Dm \sigma
\exp\Big(-\frac{N}{2g}\int d^{3}r 
[D_{\mu} \zb D_{\mu}z      
-\gamma(\zb \Jm D_{\mu}\zb)(z \Jm D_{\mu}z) 
\;\; +\sigma (|z|^{2} - 1)]
\Big), 
\label{zc}
\end{align}
where we have introduced the factor $N$ in front of the action 
to perform the $1/N$-expansion in the analytical study\cite{an}. 
At $\gamma=0$, the system (\ref{zc}) has the global SU(N) symmetry,
$z(r)\rightarrow Vz(r),\; V\in$ SU(N).
We first consider the case of small $\gamma$, and put the following 
parameterization for $z$,
$
z=z_{0}+u+iv, 
$
where 
$
z_{0}=(n_{0},0,\cdots,0), \; u=(0,u_{2},\cdots,u_{N}) \;
\mbox{and} \; v=(0,v_{2},\cdots,v_{N}). 
$
The fields $u$ and $v$ are real vectors.
We have used the gauge-fixing condition $v_1=0$. 
After substituting the above parameterization to Eq.(\ref{zc})
and keeping the quadratic terms, we perform the
Gaussian integration over $u$ and $v$ as the leading order of $1/N$,
\begin{align}
Z
&= \int \Dm n_{0} \Dm \sigma \Dm u \Dm v
\exp\Big[
-\frac{N}{2g}\int d^{3}r
\{ (\partial_{\mu}u)^{2} + (\partial_{\mu}v)^{2} 
+ \sigma(n_{0}^{2} + u^{2} + v^{2} - 1)\}
\Big] \notag\\
&= \int \Dm n_{0} \Dm \sigma \exp(-S_{\mathrm{eff}}(n_{0},\sigma)), 
\label{zd}
\end{align}
where 
\begin{equation}
S_{\mathrm{eff}}(n_{0},\sigma)
= (N-1)\mathrm{Tr}\log(-\partial_{\mu}^{2}+\sigma) 
+ \frac{N}{2g}\int d^{3}x  \sigma(n_{0}^{2} -1). 
\label{seff}
\end{equation}
As the $\gamma$-term generates only higher order terms of $u$ and $v$,
it does not give any effect in the leading order of $1/N$.

Gap equations are obtained as follows from $S_{\mathrm{eff}}(n_{0},\sigma)$ 
in Eq.(\ref{seff}),
\begin{align}
\frac{\delta S_{\mathrm{eff}}(n_{0},\sigma) }{\delta \sigma}
&= (N-1) \int \frac{d^{3}k}{(2\pi)^{3}}\frac{1}{k^{2}+ \sigma} 
+ \frac{N}{2g}(n_{0}^{2}-1)=0 
\label{gapeqa} \\
\frac{\delta S_{\mathrm{eff}}(n_{0},\sigma) }{\delta n_{0}}
&= \frac{N}{g}\; \sigma n_{0}=0. 
\label{gapeqb}
\end{align}
We use the Pauli-Villars regularization with a cutoff $\Lambda \;(\sim \pi/a)$
for the integral (\ref{gapeqa}),
and obtain the critical coupling $g_c$ by putting $\sigma=n_0=0$,
\begin{equation}
{1 \over g_c}={1 \over 2\pi}\Lambda.
\label{gc}
\end{equation}
There are two phases, one for $g>g_c$,
\begin{align}
\sqrt{\sigma_0} = \frac{2\pi}{N}\left(\frac{1}{g_{c}}-\frac{1}{g}\right), 
\;\; n_0=0, \;\; \sigma_0=\langle \sigma \rangle,
\label{sola}
\end{align}
and the other for $g<g_c$,
\begin{equation}
n_{0}^{2} = 1 - \frac{g}{g_{c}}, \;\; \sigma_0=0. 
\label{solb}
\end{equation}
For $g<g_c$, Sp(N) symmetry is spontaneously broken and 
\begin{eqnarray}
&&\sum_{a,b}\langle (\bar{z}(r)\Gamma^{ab}z(r))(\bar{z}(r')\Gamma^{ab}z(r'))\rangle
\neq 0, \;\; \mbox{for} \; |r-r'|\rightarrow \infty, \nonumber  \\
&&\sum_{a}\langle (\bar{z}(r)\Gamma^{a}z(r))(\bar{z}(r')\Gamma^{a}z(r'))\rangle
\neq 0, \;\; \mbox{for} \; |r-r'|\rightarrow \infty.
\label{corG}
\end{eqnarray}
In later section, the above result will be verified by the numerical study
of the lattice model for the effective field theory\cite{fnx}.

\subsection{Case $\gamma \approx 1$: Auxiliary fields}
In this subsection, we shall consider the case $\gamma \approx 1$.
It is useful to introduce two kinds of auxiliary vector fields $\lambda_\mu$
and $\omega_\mu$ to investigate the phase structure of the model.
Inserting the following identities to the partition function
(\ref{zc}) (where $Z_1, Z_2$ are irrelevant normalization constants and will be
ignored hereafter),
\begin{align}
1 &= Z_{1}\int \Dm \lambda_{\mu} 
\exp\left[-\frac{N}{2g}\int d^{3}r(\lambda_{\mu}-i\zb \partial_{\mu}z )^{2} 
\right]
\label{eqde} \\
1 &= Z_{2}\int \Dm \bar{\omega} D \omega
\exp\left[-\frac{N\gamma}{2g}\int d^{3}r|\omega_{\mu}-z\Jm D_{\mu}z |^{2} 
\right],
\label{identitiy}
\end{align}
we obtain,
\begin{align}
Z &= \int \Dm z\Dm\zb\Dm\sigma\Dm\lambda_{\mu}\Dm\omegab_{\mu}\Dm\omega_{\mu}
\exp\bigg(
-\frac{N}{2g}\int d^{3}r [\zb(-\partial_{\mu}^{2}
+i\lambda_{\mu}\overleftrightarrow{\partial}_{\mu}+\sigma)z \notag\\
&\;\;\; -z  (\gamma \omegab_{\mu}\Jm\partial_{\mu})z
-\zb(\gamma \omega_{\mu}\Jm\partial_{\mu})\zb
+\lambda_{\mu}^{2}+\gamma|\omega_{\mu}|^{2} -\sigma ]
\bigg) \notag\\
&= \int \Dm z\Dm\zb\Dm\sigma\Dm\lambda_{\mu}\Dm\omegab_{\mu}\Dm\omega_{\mu}
 \exp\bigg(
-\frac{N}{2g}\int d^{3}r [\zb(-\partial_{\mu}^{2}
+i\lambda_{\mu}\overleftrightarrow{\partial}_{\mu}+\lambda_{\mu}^{2} \notag\\
&\;\;\; +\gamma|\omega_{\mu}|^{2}+\sigma)z-z  
(\gamma \omegab_{\mu}\Jm\partial_{\mu})z
-\zb(\gamma \omega_{\mu}\Jm\partial_{\mu})\zb
 -\sigma ]
\bigg),
\label{ze}
\end{align}
where $\bar{z}\lambda_\mu \overleftrightarrow{\partial}_{\mu}z
=\lambda_\mu(\bar{z}\cdot\partial_\mu z -\partial_\mu\bar{z}\cdot z)$.

\subsubsection{Strong-coupling region}

First we shall study the model (\ref{ze}) in the strong-coupling
region $g>g_c$, in which $n_0=0,\; \langle \sigma \rangle>0$.
We put $B = \gamma \omega_{\mu}\Jm\partial_{\mu}, 
\ \bar{B}=\gamma \omegab_{\mu}\Jm\partial_{\mu}$ for notational simplicity.
Then the partition function is expressed as 
\begin{align}
Z &= 
\int \Dm z\Dm\zb\Dm\sigma\Dm\lambda_{\mu}\Dm\omegab_{\mu}\Dm\omega_{\mu}
\exp\bigg(
-\frac{N}{2g}\int d^{3}r [
u(-\partial_{\mu}^{2}+\lambda_{\mu}^{2} \notag\\
&+\sigma+\gamma|\omega_{\mu}|^{2}+B+\bar{B})u 
+u\{-\lambda_{\mu}\overleftrightarrow{\partial}_{\mu}+i(\bar{B}-B)\}v \notag\\
&+v\{ \lambda_{\mu}\overleftrightarrow{\partial}_{\mu}+i(\bar{B}-B)\}u 
+v\{-\partial_{\mu}^{2}+\lambda_{\mu}^{2}+\sigma
+\gamma|\omega_{\mu}|^{2} \notag\\
&-(B+\bar{B})\}v-\sigma
 ]
\bigg),
\label{zf}
\end{align}
where $z=u+iv$.
As the action in Eq.(\ref{zf}) is a quadratic form of $z=u+iv$, 
integration over $z$ can be done.
We define
\begin{align}
\alpha &= -\partial_{\mu}^{2} +\lambda_{\mu}^{2}
+\gamma|\omega_{\mu}|^{2}+ \sigma, \notag\\
\beta  
&=\sqrt{(\lambda_{\mu}\overleftrightarrow{\partial}_{\mu})^{2}
+\gamma^{2}|\omega_{\mu}\dellr_{\mu}|^{2}},
\end{align}
and the result after the integration over $z$ is given as 
\begin{align}
Z=\int \Dm\sigma\Dm\lambda_{\mu}\Dm\omegab_{\mu}\Dm\omega_{\mu}
\exp\Big(-\frac{N}{2}\mathrm{Tr}\log(\alpha^{2}+\beta^{2}) 
+\frac{N}{2g}\int dx^{3}\sigma
\Big),
\label{zg}
\end{align}
where
\begin{align}
\log(\alpha^{2}+\beta^{2})
=\log\Big\{(-\partial_{\mu}^{2} +\lambda_{\mu}^{2}
+\gamma|\omega_{\mu}|^{2}+ \sigma)^{2}+
 (\lambda_{\mu}\overleftrightarrow{\partial}_{\mu})^{2} 
+\gamma^{2}|\omega_{\mu}\dellr_{\mu}|^{2}\Big\}.\nonumber
\end{align}

From Eq.(\ref{zg}), we shall obtain an effective action of the 
vector fields $\lambda_\mu$ and $\omega_\mu$ in power of them.
To this end, we use the identity
\begin{align}
&\log(\alpha^{2}+\beta^{2})
=2\log(-\partial_{\mu}^{2}+\sigma) \notag\\
&\hspace{1cm}+\Tr\bigg[\frac{2}{-\partial_{\mu}^{2}+\sigma}\lambda_{\mu}^{2} 
+\frac{1}{(-\partial_{\mu}^{2}+\sigma)^{2}}
(\lambda_{\mu}\dellr_{\mu})^{2}\bigg] \notag\\
&\hspace{1cm}+\Tr\left[\frac{2\gamma}{-\partial_{\mu}^{2}+\sigma}|\omega_{\mu}|^{2}
+\frac{\gamma^{2}}{(-\partial_{\mu}^{2}+\sigma)^{2}}
|\omega_{\mu}\dellr_{\mu}|^{2}\right] \notag\\
&\hspace{1cm} +O(\lambda^4_\mu, \omega^4_\mu, \lambda^2_\mu \omega^2_\mu),
\label{logab}
\end{align}
for positive constant $\sigma$.
We explicitly evaluate the momentum integrals in Eq.(\ref{logab}),
\begin{align}
&\Tr\left[\frac{2}{-\partial_{\mu}^{2}+\sigma}\lambda_{\mu}^{2}
+\frac{1}{(-\partial_{\mu}^{2}+\sigma)^{2}}
(\lambda_{\mu}\dellr_{\mu})^{2}\right] \notag\\
&\hspace{1cm}=\int \frac{d^{3}p}{(2\pi)^{3}}
\lambda_{\mu}(p)\Pi_{\mu\nu}(p)\lambda_{\nu}(-p),
\label{eqea}
\end{align}
\begin{align}
\Tr\left[\frac{2\gamma}{-\partial_{\mu}^{2}
+\sigma}|\omega_{\mu}|^{2}\right]
=\gamma\int \frac{d^{3}p}{(2\pi)^{3}}
\omegab_{\mu}(p)\Omega^{(1)}_{\mu\nu}\omega_{\nu}(p),
\label{eqeb}
\end{align}
\begin{align}
\Tr\left[\frac{\gamma^{2}}{(-\partial_{\mu}^{2}+\sigma)^{2}}
|\omega_{\mu}\dellr_{\mu}|^{2}\right]
=\gamma^{2}\int \frac{d^{3}p}{(2\pi)^{3}}
\omegab_{\mu}(p)\Omega^{(2)}_{\mu\nu}(p)\omega_{\nu}(p),
\label{eqec}
\end{align}
where
\begin{align}
\Pi_{\mu\nu}(p)
&=\int \frac{d^{3}q}{(2\pi)^{3}}
\left[\frac{\delta_{\mu\nu}}{q^{2}+\sigma}
-\frac{(p+2q)_{\mu}(p+2q)_{\nu}}{2((p+q)^{2}+\sigma)
(q^{2}+\sigma)}\right] \notag\\
&=\left(\delta_{\mu\nu}-\frac{p_{\mu}p_{\nu}}{p^{2}}\right)\Pi(p), 
\label{eqed}\\
\Pi(p)
&=\frac{1}{2}(p^{2}+4\sigma)\Sigma(p)-\frac{\sqrt{\sigma}}{4\pi}, 
\hspace{12pt} \notag\\
\Sigma(p)
&=\frac{1}{4\pi|p|}\tan^{-1}\sqrt{\frac{p^{2}}{4\sigma}}, \notag\\
\Omega^{(1)}_{\mu\nu} 
&=\int\frac{d^{3}q}{(2\pi)^{3}}\frac{\delta_{\mu\nu}}{q^{2}+\sigma}
=\Omega^{(1)}\delta_{\mu\nu}, \notag\\
\Omega^{(2)}_{\mu\nu}(p) 
&=-\int \frac{d^{3}q}{(2\pi)^{3}}
\frac{(p+2q)_{\mu}(p+2q)_{\nu}}{2((p+q)^{2}+\sigma)(q^{2}+\sigma)}. 
\label{eqeg}
\end{align}
Then effective action of the vector fields $\lambda_\mu$ and
$\omega_\mu$ is obtained as follows up to the quadratic order of them,
\begin{align}
S^{(2)}_{\rm eff}(\lambda_\mu,\omega_\mu)
&=N\int \frac{d^{3}p}{(2\pi)^{3}}\lambda_{\mu}(p)
\Pi_{\mu\nu}(p)\lambda_{\nu}(-p)\notag\\
&+N\gamma^{2}\int \frac{d^{3}p}{(2\pi)^{3}}
\omegab_{\mu}(p)\Pi_{\mu\nu}(p)\omega_{\nu}(p) \notag\\
&+N\gamma(1-\gamma)\int \frac{d^{3}p}{(2\pi)^{3}}
\omegab_{\mu}(p)\Omega^{(1)}\omega_{\nu}(p) \notag\\
&-\frac{N}{2g}\int\frac{d^{3}p}{(2\pi)^{3}} \sigma.
\label{seffb}
\end{align}
For positive $\sigma$ and $p \ll 1$,
\begin{align}
\Sigma(p) &\approx \frac{1}{4\pi p}\sqrt{\frac{p^{2}}{4\sigma}} \label{eqeh}\\
\Pi(p) &\approx \frac{1}{2}(p^{2}+4\sigma)\frac{1}{8\pi\sqrt{\sigma}}
-\frac{\sqrt{\sigma}}{4\pi}
=\frac{p^{2}}{16\pi\sqrt{\sigma}} \label{eq:59}\\
\Pi_{\mu\nu}(p) &\approx \frac{1}{16\pi\sqrt{\sigma}}
(p^{2}\delta_{\mu\nu}-p_{\mu}p_{\nu}), \label{eq:60}
\end{align}
and therefore the quadratic term of $\omega_\mu$ becomes as follows at
low momentum,
\begin{align}
&N\int \frac{d^{3}p}{(2\pi)^{3}}
\omegab_{\mu}(p)
\Big[\frac{\gamma^{2}}{16\pi\sqrt{\sigma}}
(p^{2}\delta_{\mu\nu}-p_{\mu}p_{\nu}) 
+\gamma(1-\gamma)\Omega^{(1)}\delta_{\mu\nu}
\Big]\omega_{\nu}(p). 
\label{omega2}
\end{align}
From Eq.(\ref{omega2}), it is obvious that $\omega_\mu$ behaves
like a massive vector field for $\gamma<1$, for $\gamma=1$
it becomes massless and behaves like a kind of gauge field, 
and finally for $\gamma>1$ its
nonvanishing condensation is expected to occur.
More detailed study on the case $\gamma=1$ will be given in later
section, and it is shown there that a SU(2) gauge model really appears.

Let us study the case $\gamma>1$ somewhat in detail for the large-$N$ limit.
Condensation of $\omega_\mu$ apparently breaks the rotational
symmetry of the space (or $\frac{\pi}{2}$-rotation of the square lattice)
and also the U(1) gauge symmetry to $Z_2$\cite{nakane}.
Here we assume $\langle \omega_x \rangle=\omega\neq 0$
without loss of generality.
We also assume that $\omega$ is real by the gauge symmetry of the
system.
Then the action of $z(r)$ is given as 
\begin{align}
S_z &= \frac{N}{2g} \int d^{3}r
[-\zb \partial_{\mu}^{2}z -\gamma\omega(\zb\Jm\partial_{x}\zb)
-\gamma{\omega}(\zb\Jm\partial_{x}\zb) 
+\sigma(|z|^2-1)].
\label{sz}
\end{align}
The action $S_z$ in Eq.(\ref{sz}) can be diagonalized by 
introducing field $\xi(r)$ as
\begin{equation}
z(r) = \frac{1}{\sqrt{2}}
\{e^{-i\gamma \omega x}\xi(r)
+e^{i\gamma \omega  x}(i\Jm\bar{\xi}(r)) \},
\label{xi}
\end{equation}
\begin{align}
S_\xi&=S_z 
= \frac{N}{2g}\int d^{3}r
\{\bar{\xi}(-\partial_{\mu}^{2} + \gamma(1-\gamma)\omega^{2})\xi + 
\sigma(\bar{\xi}\xi -1) \}.
\label{sxi}
\end{align}
From Eq.(\ref{sxi}), it is obvious that the field $\xi(r)$ acquires 
its mass squared $\sigma' =\sigma + \gamma(1-\gamma)\omega^{2}$. 

From $S_\xi$ in Eq.(\ref{sxi}), we can have a gap equation and determine
the critical coupling $g_c$.
By integrating out $\xi(r)$, we obtain 
\begin{align}
S'_{\mathrm{eff}}(\sigma') = 
N\mathrm{Tr}\log(-\partial_{\mu}^{2}+ \sigma') 
- \frac{N}{2g}\int d^{3}x 
\{\sigma' - \gamma(1-\gamma)\omega^{2} \}, 
\label{s'eff}
\end{align}
and 
\begin{equation}
\frac{S'_{\mathrm{eff}}(\sigma')}{\delta \sigma'}
=N \int \frac{d^{3}k}{(2\pi)^{3}}\frac{1}{k^{2}+ \sigma'} 
- \frac{N}{2g}=0. 
\label{gapeqc}
\end{equation}
Solution to Eq.(\ref{gapeqc}) is obtained as 
\begin{align}
\sqrt{\sigma'} &= 2\pi\left(\frac{1}{g_{c}} - \frac{1}{g}\right) 
\label{sol2_1} \\
\frac{1}{g_{c}} &= \frac{\Lambda}{2\pi}. 
\label{solbb}
\end{align}
The above critical value $g_c$ is the same with that obtained for the case
of small $\gamma$.

\begin{figure}[th]
\begin{center}
\includegraphics[width=5cm]{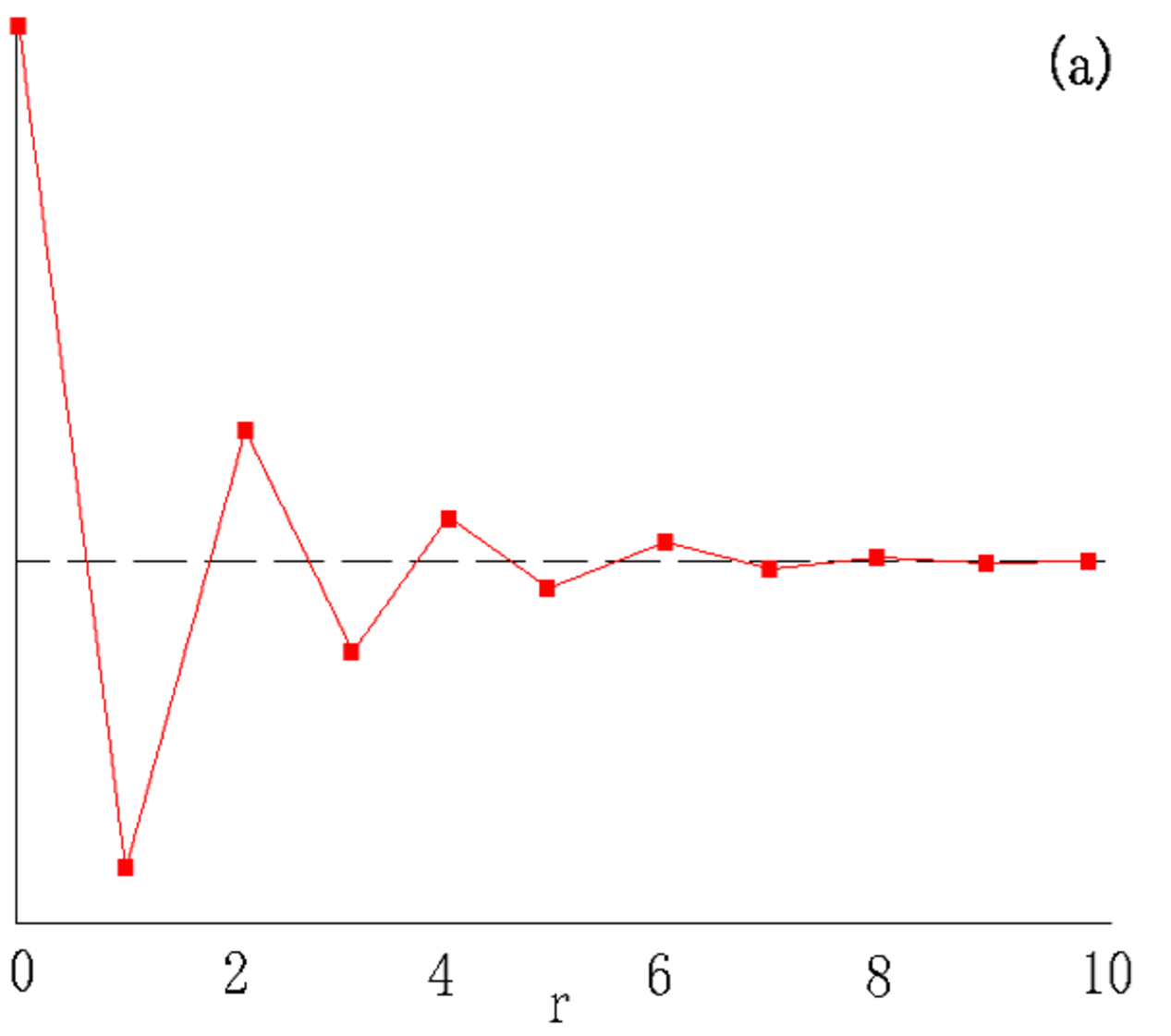}
\includegraphics[width=5cm]{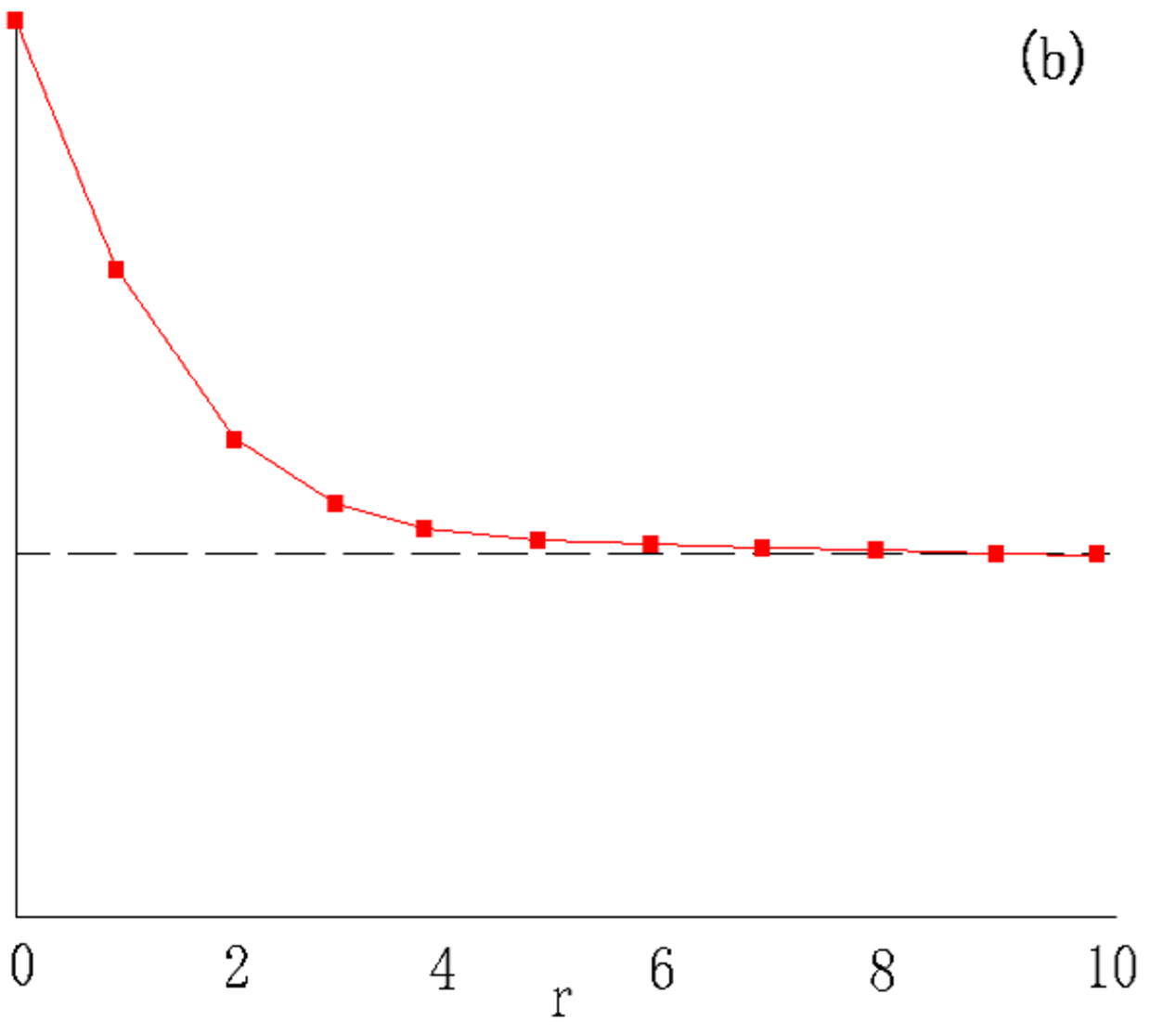}
\end{center}
\caption{(a)Correlation of $\Gamma^{ab}$
(b)Correlation of $\Gamma^a$.}
\label{srspiral}
\end{figure}

In the large-$N$ limit, spin correlations are obtained from
Eq.(\ref{xi}), see Fig.\ref{srspiral}.


\subsubsection{Weak-coupling region}
We shall consider the weak coupling region $g<g_c$, in which the
condensation of $z(r)$ occurs.
We assume a smooth second-order phase transition from the phase of
$\omega=0$ to that of $\omega\neq 0$
as $\gamma$ is increased,
and estimate the critical coupling $\gamma_c$\cite{fnb}.
From Eq.(\ref{xi}), we divide $z$ as $z=z_{0}+\tilde{z}$, 
where $z_{0} = (z_{1},0, \cdots,0,z_{N})$, 
$\tilde{z} = (0,z_{2},\cdots,z_{N-1},0)$.
Furthermore we put $\langle \sigma \rangle=0$.
Substituting the above expression of $z$ into the action,
we obtain
\begin{align}
S &= \frac{N}{2g}
\int d^{3}r \Big[\zb(-\partial_{\mu}^{2}+\lambda_{\mu}^{2}
+\gamma|\omega_{\mu}|^{2}+\sigma)z \notag\\
&+i\lambda_{\mu}\zb\dellr_{\mu} z 
-\bar{\omega}_\mu z\Jm\partial_{\mu}z 
- \omega_\mu \zb\Jm\partial_{\mu}\zb -\sigma\Big] \notag\\
&=\tilde{S} + S_{0}, 
\label{ssx}
\end{align}
where
\begin{align}
\tilde{S} 
&= \frac{N}{2g}\int d^{3}r \Big[\bar{\tilde{z}}
(-\partial_{\mu}^{2}+\lambda_{\mu}^{2}
+\gamma|\omega_{\mu}|^{2}+\sigma)\tilde{z} \notag\\
&+i\lambda_{\mu}\bar{\tilde{z}}\dellr_{\mu}\tilde{z} 
- \bar{\omega}_\mu\tilde{z}\Jm\partial_{\mu}\tilde{z}
- \omega_\mu \bar{\tilde{z}}\Jm\partial_{\mu}\bar{\tilde{z}} -\sigma\Big] 
\label{tilS}\\
S_{0}
&= \frac{N}{2g}\int d^{3}r \Big[\zb_{0}(-\partial_{\mu}^{2}+\lambda_{\mu}^{2}
+\gamma|\omega_{\mu}|^{2}+\sigma)z_{0} \notag\\
&+i\lambda_{\mu}\zb_{0}\dellr_{\mu} z_{0} 
- \bar{\omega}_\mu z_{0}\Jm\partial_{\mu}z_{0} - \omega_\mu \zb_{0}
\Jm\partial_{\mu}\zb_{0} \Big]. 
\label{sx}
\end{align}
Integration over $\tilde{z}_x$ can be performed as in the previous case,
\begin{align}
&\int \Dm \bar{\tilde{z}}\Dm \tilde{z} \exp(-\tilde{S}) \notag\\
&\hspace{0.5cm} = \exp \Big[-\frac{N-4}{2}\int d^3r
\Big\{(-\partial^2_{\mu} +\lambda^2_{\mu}
+\gamma|\omega_{\mu}|^{2}+ \sigma)^{2}  \notag\\
&\hspace{1cm} +(\lambda_{\mu}\overleftrightarrow{\partial}_{\mu})^2  
+\gamma^{2}|\omega_{\mu}\partial_{\mu}|^2\Big\}  
+{N \over 2g} \int d^3r \;  \sigma \Big].
\label{inttilz}
\end{align}
For $S_0$, we substitute the following expression for $z_0$,
\begin{align}
z_{0} &= N_{0} + z',\; 
N_{0} = (\frac{1}{\sqrt{2}}n_{0},0, \cdots,0,\frac{1}{\sqrt{2}}n_{0}), \notag\\
z' &= (z_{1},0, \cdots,0,z_{N}), 
\label{zs}
\end{align}
and obtain
\begin{align}
S_{0} &= \frac{N}{2g}\int d^{3}r 
\Big[\zb'(-\partial_{\mu}^{2}+\lambda_{\mu}^{2}
+\gamma|\omega_{\mu}|^{2}+\sigma)z' \notag\\
&+i\lambda_{\mu}\zb'\dellr_{\mu}z' 
- \bar{\omega}_\mu z'\Jm\partial_{\mu}z' 
- \omega_\mu \zb'\Jm\partial_{\mu}\zb'\Big] \notag\\
&+ S_{\mathrm{int}},
\label{eqgg}
\end{align}
where
\begin{align}
S_{\mathrm{int}} 
&= \frac{N}{2g}\int d^{3}r
\Big[(\lambda_{\mu}^{2}+\gamma|\omega_{\mu}|^{2}+\sigma)
\{N_{0}^{2}+N_{0}(\zb'+z')\}  \notag\\
&+i\lambda_{\mu}N_{0}\partial_{\mu}
(z'-\zb') 
-\gamma\bar{\omega}_{\mu}N_{0}\Jm\partial_{\mu}z'
-\gamma\omega_{\mu}N_{0}\Jm\partial_{\mu}\zb'\Big]. 
\label{eqgh}
\end{align}
We cannot integrate out $z'$ exactly and therefore
treat the cubic terms in $S_{\mathrm{int}}$ as a ``perturbation" of the
$1/N$-expansion\cite{an}.
Then the final expression of the effective action $S'_{\rm eff}(\lambda_\mu,\omega_\mu)$ 
is given as follows up the quadratic terms of 
$\lambda_\mu,\; \sigma$ and $\omega_\mu$,
\begin{align}
S'_{\rm eff}(\lambda_\mu,\omega_\mu) &=
{N}\lambda_{\mu}\left\{\Pi_{\mu\nu}+n_{0}^{2}\left(\delta_{\mu\nu} 
- \frac{p_{\mu}p_{\nu}}{p^{2}}\right)\right\}\lambda_{\mu} \notag\\
&+{N}\gamma^{2}\omega_{\mu}^{R}
\Big\{\Pi_{\mu\nu}+n_{0}^{2}\Big(\frac{1}{\gamma}
\delta_{\mu\nu} 
- \frac{p_{\mu}p_{\nu}}{p^{2}}\Big) 
+ \frac{1-\gamma}{\gamma}\Omega_{\mu\nu}^{(1)}\Big\}
\omega_{\mu}^{R} \notag\\
&+{N}\gamma^{2}\omega_{\mu}^{I}\left\{\Pi_{\mu\nu}+n_{0}^{2}
\left(\frac{1}{\gamma}
\delta_{\mu\nu} 
- \frac{p_{\mu}p_{\nu}}{p^{2}}\right) + 
\frac{1-\gamma}{\gamma}\Omega_{\mu\nu}^{(1)}\right\}
\omega_{\mu}^{I} \notag\\
&+{N}\sigma\left( \frac{1}{2}\Sigma(p) + n^{2}_{0}\right)\sigma
- \frac{N}{2g}\; \sigma, 
\label{eqgi}
\end{align}
where $\omega_{\mu} = \omega_{\mu}^{R}+i\omega_{\mu}^{I}$.

It is obvious that for $n_0\neq 0$ case, only the transverse modes of
$\omega_\mu$ (i.e., $\sum_\mu p_\mu\omega_\mu=0$) can survive at low momentum.
Furthermore its mass is renormalized by $n_0\neq 0$ as 
$\gamma[n^2_0+(1-\gamma)\Omega^{(1)}]$.
Then for $\gamma > \gamma_c=1+{n^2_0(g) \over \Omega^{(1)}}$, $\xi(r)$, 
instead of
$z(r)$, condenses and nontrivial correlations of the ``spin operators",
$\hat{\Gamma}^{ab}$, appear as a result of 
$\langle \omega_\mu \rangle \neq 0$.
In this phase, 
\begin{eqnarray}
\langle\hat{\Gamma}^{ab}(r)\rangle
&=&\langle\bar{z}(r)\Gamma^{ab}z(r)\rangle \nonumber \\
&=&n_1^{ab}\cos (2\gamma \omega x)+n_2^{ab}\sin (2\gamma \omega x),
\label{spiralspina}
\end{eqnarray}
where $n_1^{ab}=\mbox{Re}[\langle\xi\rangle\Jm\Gamma^{ab}\langle\xi\rangle],\;
n_2^{ab}=\mbox{Im}[\langle\xi\rangle\Jm\Gamma^{ab}\langle\xi\rangle]$.
On the other hand, 
$\langle\hat{\Gamma}^{a}(r)\rangle=\langle\bar{\xi}\rangle
\Gamma^{a}\langle\xi\rangle$.

\section{SU(2) gauge theory at $\gamma=1$}
\setcounter{equation}{0}

In the previous section, we found that the composite vector field
$\omega_\mu$ behaves like a massless gauge field at $\gamma=1$,
and for $g<g_c$ it acquires mass squared propotional to 
$n_0^2=\langle \bar{z}\cdot z \rangle$ as a result of the 
Higgs mechanism. 
In this section, we shall explicitly show that the three real
vector fields $(\lambda_\mu, \omega^R_\mu, \omega^I_\mu)$ form
a SU(2) gauge field minimally coupled with $z_x$ at $\gamma=1$.
This is another example of the symmetry-enhancement phenomenon,
i.e., emergent symmetry at low energies.

We start with the action $S_{\gamma=1}$ in Eq.(\ref{ze}),
\begin{align}
S_{\gamma=1}&=
\frac{N}{2g} \int d^3r \Big[\zb(-\partial_{\mu}^{2}
+i\lambda_{\mu}\overleftrightarrow{\partial}_{\mu}+\lambda_{\mu}^{2} 
+|\omega_{\mu}|^{2}+\sigma)z  
-z(\omegab_{\mu}\Jm\partial_{\mu})z
-\zb(\omega_{\mu}\Jm\partial_{\mu})\zb
 -\sigma \Big],
\label{lg}
\end{align}
where we have put $\gamma=1$.
Hereafter we explicitly consider the CP$^3$ case but generalization
to an arbitrary $N$ is straightforward.
We first redefine the CP$^{3}$ field $Z(r)$ from the original 
$z(r)$ as follows,
$Z(r)=(z_{1}(r),z_{2}(r),\bar{z}_{4}(r),-\bar{z}_{3}(r))^t$.
It is easily verified that $Z(r)$ is a CP$^3$ field, 
$\sum^4_{i=1}|Z_{i}(r)|^2=1.$

It is straightfward to verify the following equation,
\begin{equation}
\bar{z}\dellr_\mu z=\bar{Z}\Sigma_3\dellr_\mu Z,
\label{zidenta}
\end{equation}
where 
\begin{equation}
\Sigma_3
= \left( \begin{array}{cccc}
1 & 0  & 0 & 0 \\
0 & 1 & 0 & 0 \\
0 & 0 & -1 & 0 \\
0 & 0 & 0 & -1 
\end{array}
\right).
\label{sigmac}
\end{equation}
Similarly
\begin{align}
\bar{\omega}_\mu (z\Jm\partial_{\mu})z
+\omega_{\mu}(\zb\Jm\partial_{\mu})\zb 
=i\omega^R_\mu (\bar{Z}\Sigma_2\dellr_\mu Z)
+i\omega^I_\mu (\bar{Z}\Sigma_1\dellr_\mu Z),
\label{zidentb}
\end{align}
where
\begin{equation}
\Jm
= \left( \begin{array}{cccc}
0 & 0  & 0 & 1 \\
0 & 0 & -1 & 0 \\
0 & 1 & 0 & 0 \\
-1 & 0 & 0 & 0
\end{array}
\right),
\;\;
\Sigma_2
= \left( \begin{array}{cccc}
0 & 0  & i & 0 \\
0 & 0 & 0 & i \\
-i & 0 & 0 & 0 \\
0 & -i & 0 & 0
\end{array}
\right),
\nonumber  
\end{equation}
\begin{equation}
\Sigma_1
= \left( \begin{array}{cccc}
0 & 0  & 1 & 0 \\
0 & 0 & 0 & 1 \\
1 & 0 & 0 & 0 \\
0 & 1 & 0 & 0
\end{array}
\right).
\end{equation}
It is obvious that $\Sigma_i(i=1,2,3)$ satisfy the SU(2) algebra.
Let us define SU(2) gauge field $\vec{\cal A}_\mu$ as 
$\vec{\cal A}_\mu=(\lambda^1_\mu, \lambda^2_\mu, \lambda^3_\mu)
=(\omega^I_\mu, \omega^R_\mu,\lambda_\mu)$, then the Lagrangian (\ref{lg})
can be rewritten as follows,
\begin{equation}
S_{\gamma=1}=
\frac{N}{2g}\int d^3r
\Big[|(\partial_\mu+i\vec{\Sigma}\cdot\vec{\cal A}_\mu)Z|^2
+\sigma |Z|^2-\sigma\Big].
\label{lgsub}
\end{equation}

\begin{figure}[ht]
\begin{center}
\includegraphics[width=5cm]{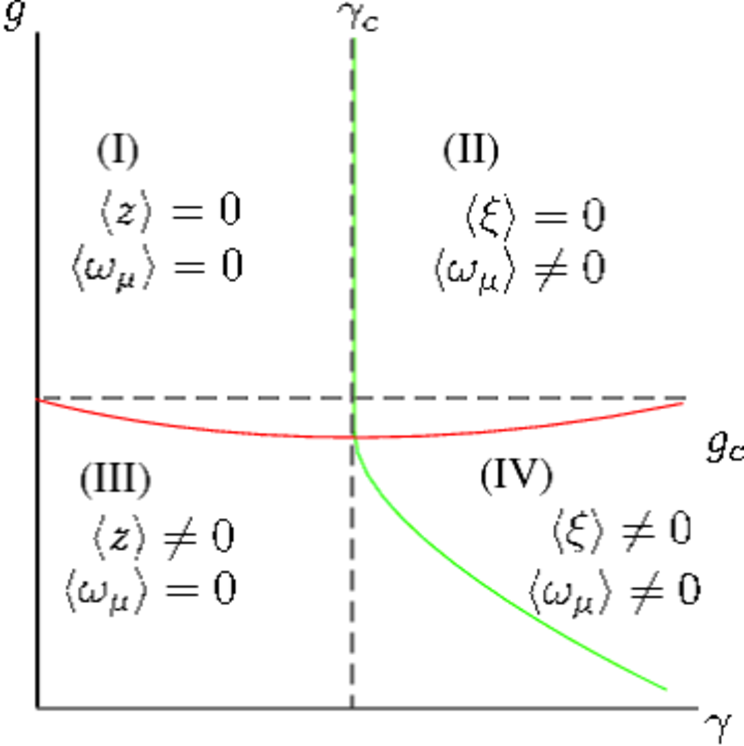}
\includegraphics[width=5cm]{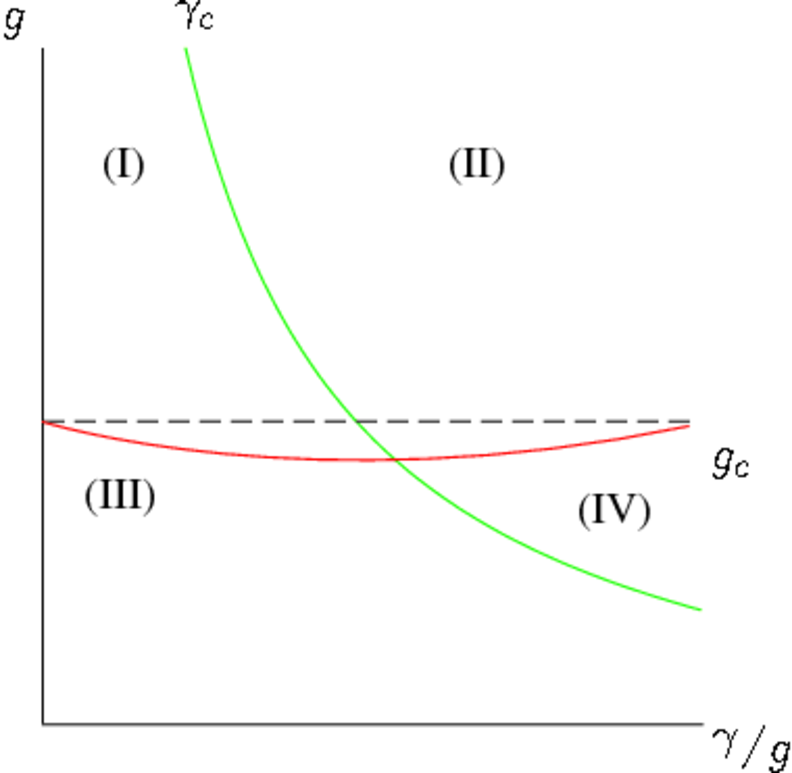}
\end{center}
\caption{Phase diagram in $(\gamma-g)$ and $(\gamma/g-g)-$ planes.}
\label{phasean}
\end{figure}

Lattice gauge model corresponding to the above SU(2) gauge theory 
(\ref{lgsub}) is under study and result will be reported in 
a future publication.
However phase structure of the system can be inferred by qualitative
discussion.
As in the usual CP$^{N-1}$ model coupled with the U(1) gauge field, 
there exists a phase transition that separates ordered and disordered phases.
However, as the SU(2) gauge field fluctuates the hopping of the spinon 
$Z(r)$ more strongly than the U(1) gauge field, the critical coupling 
$g_c(\gamma=1)$ is expected to be  smaller than $g_c(\gamma=0)$.
From this consideration, we expect that the critical coupling
$g_c(\gamma)$ is a decreasing function of $\gamma$ for 
$\gamma<\gamma_c$, though in the
previous discussion for small $\gamma\ll 1$ by the $1/N$-expansion
we do not find any $\gamma$ dependence of $g_c$.
As $\gamma$ exceeds $\gamma_c$, the condensation of $\omega_\mu$
tends to occur and fluctuations of $\omega_\mu$ are suppressed.
Moreover the original U(1) gauge symmetry reduces to $Z_2$
gauge symmetry by the Anderson-Higgs mechanism and fluctuations of 
$\lambda_\mu$ are also suppressed.
Then $g_c(\gamma)$ starts to increase at $\gamma=\gamma_c$.
The above expectation will be confirmed by the numerical study of lattice-gauge
model in the following section.
Expected phase diagram is show in the $(\gamma-g)-$ plane
in Fig.\ref{phasean}.

In the following section, we shall introduce a lattice model
for the effective field theory with general value of $\gamma$,
and study it by means of MC simulations.

\begin{figure}[th]
\begin{center}
\includegraphics[width=7cm]{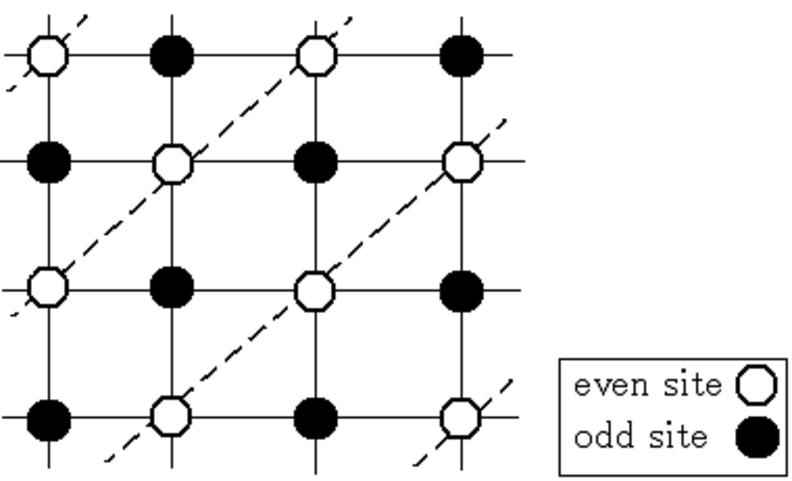}
\end{center}
\caption{Interaction terms in $A_L$ in Eq.(\ref{sl}).}
\label{aaaa}
\end{figure}
\begin{figure}[th]
\begin{center}
\includegraphics[width=8cm]{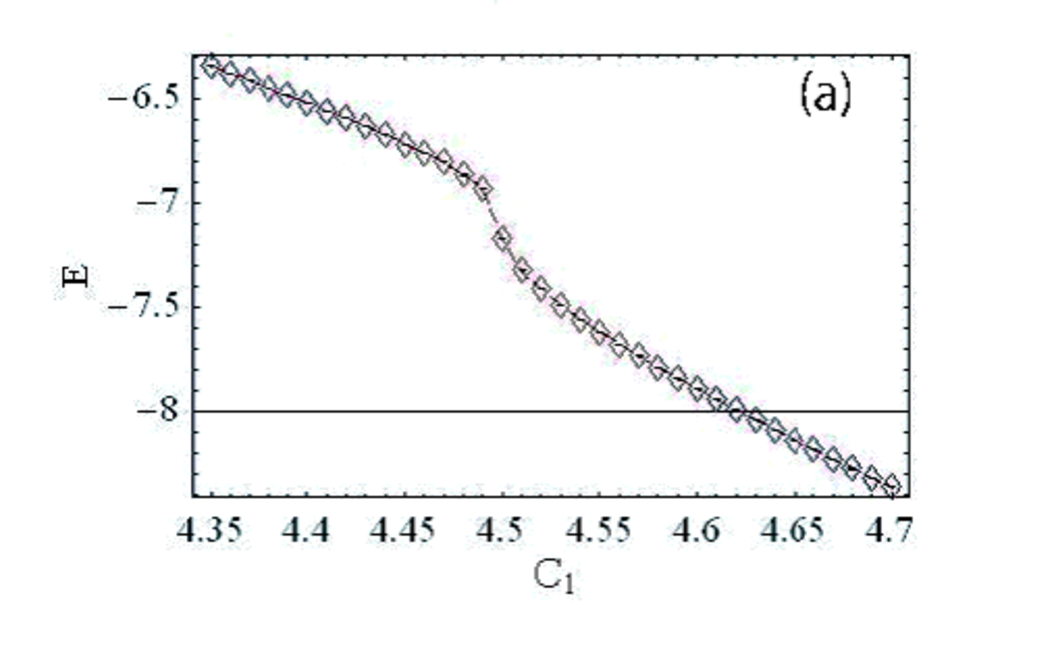}
\includegraphics[width=8cm]{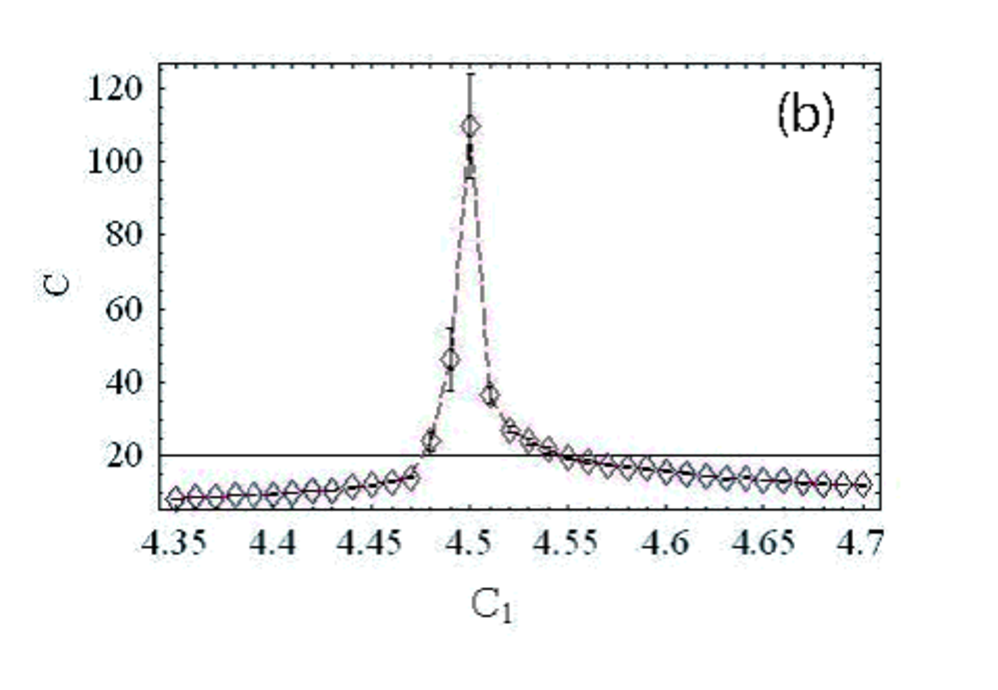}
\end{center}
\vspace{-1cm}
\caption{(a)Enegry $E$ as a function of $c_1$ for $c_2=c_3=0$.
There exists a discontinuity at $c_1=4.5$.
(b)``Specific heat" $C$ as a function of $c_1$ for $c_2=c_3=0$.
There exists a sharp peak at $c_1=4.5$.
System size $L=20$.}
\label{eb}
\end{figure}

\section{Numerical study}
\setcounter{equation}{0}
\label{sec:Lattice model}

\begin{figure}[th]
\begin{center}
\includegraphics[width=8cm]{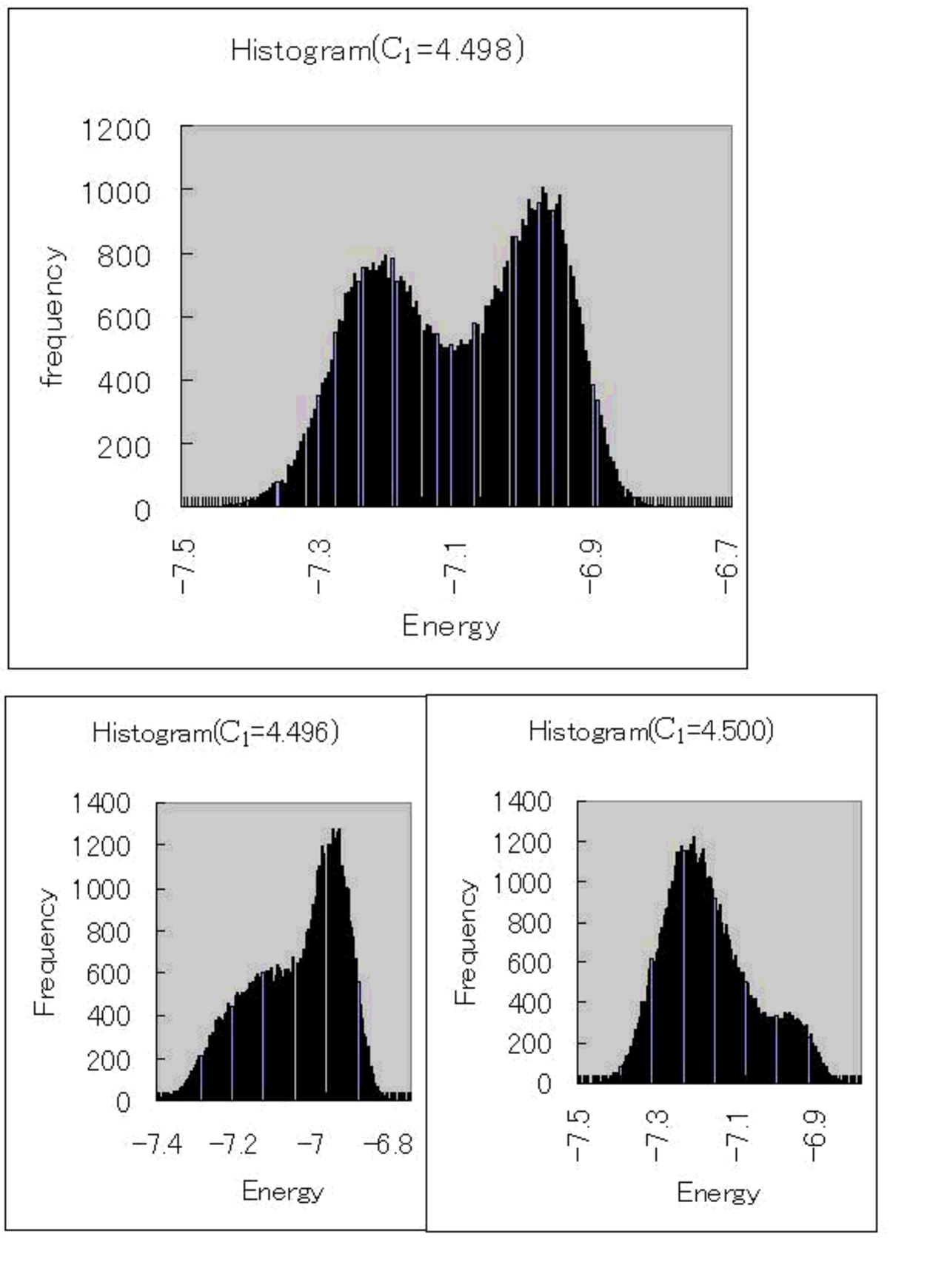}
\end{center}
\caption{Enegry distribution $N[E]$ for the pure CP$^3$ model.
At $c_1=4.498$, $N(E)$ has double-peak shape, whereas it has 
single peak at $c_1=4.496$ and $4.500$.
System size $L=20$.}
\label{histoa}
\end{figure}

In this section we formulate the effective field theory (\ref{zc}) 
on a cubic lattice,
and investigate its phase structure by means of the MC simulations.
We explicitly consider the Sp(4) model.
Action of the lattice model $A_L$ is given as follows,
\begin{align}
A_L&=c_1\sum_{r,\mu}\bar{z}_{r+\mu}U_{r\mu}z_r+\mbox{c.c.} \notag\\
    &+c_2\sum_{r,\mu,\nu}U_{r\mu}U_{r+\mu,\nu}\bar{U}_{r+\nu,\mu}\bar{U}_{r\nu}
    +\mbox{c.c.} \notag\\
    &+c_3\sum_{r,\mu} |z_r \Jm z_{r+\mu}|^2
      +c_4\sum_{r,\mu} |z_r \Jm z_{r+1+2}|^2  ,
\label{sl}
\end{align}    
where $r$ denotes the cubic lattice site, $\mu=(0,1,2)$ is the direction index
and it also denotes the unit vector in the $\mu$-direction.
Field $z_r$'s are CP$^3$ variables and $U_{r\mu}$ is a U(1) gauge field 
defined on link $(r,\mu)$, $U_{r\mu}\sim e^{i\lambda_{\mu}(r)}$.
The parameters $c_1 \propto 1/g$, $c_3\propto \gamma/g$ and the $c_2$-term
is the lattice Maxwell term (the so-called Wilson term) corresponding to 
$(\partial_\mu A_\nu-\partial_\nu A_\mu)^2$ in the continuum spacetime\cite{fnb}.
We also added the $c_4$-term on the diagonal lines in the 2D layers
that represents the exchange couplings between spins
on even sites of the original lattice.
See Fig.\ref{aaaa}.
The partition function $Z_L$ is given as
\begin{equation}
Z_L=\int [DU][DzD\bar{z}]_{\rm CP}\; e^{A_L},
\label{zlattice}
\end{equation}
where $[DzD\bar{z}]_{\rm CP}$ denotes the integration over CP$^3$ variables,
and we use a specific parameterization for them in the MC simulations.
For the MC simulations,
we used the standard Metropolis algorithm of local update.
The typical statistics was $10^5$ MC steps per sample,
and the averages and errors were estimated over ten samples.
The typical acceptance ratio was about $50$\%.
We also used multi-histogram methods to obtain reliable results near the 
phase transition point\cite{mhm}.

\begin{figure}[ht]
\begin{center}
\includegraphics[width=8cm]{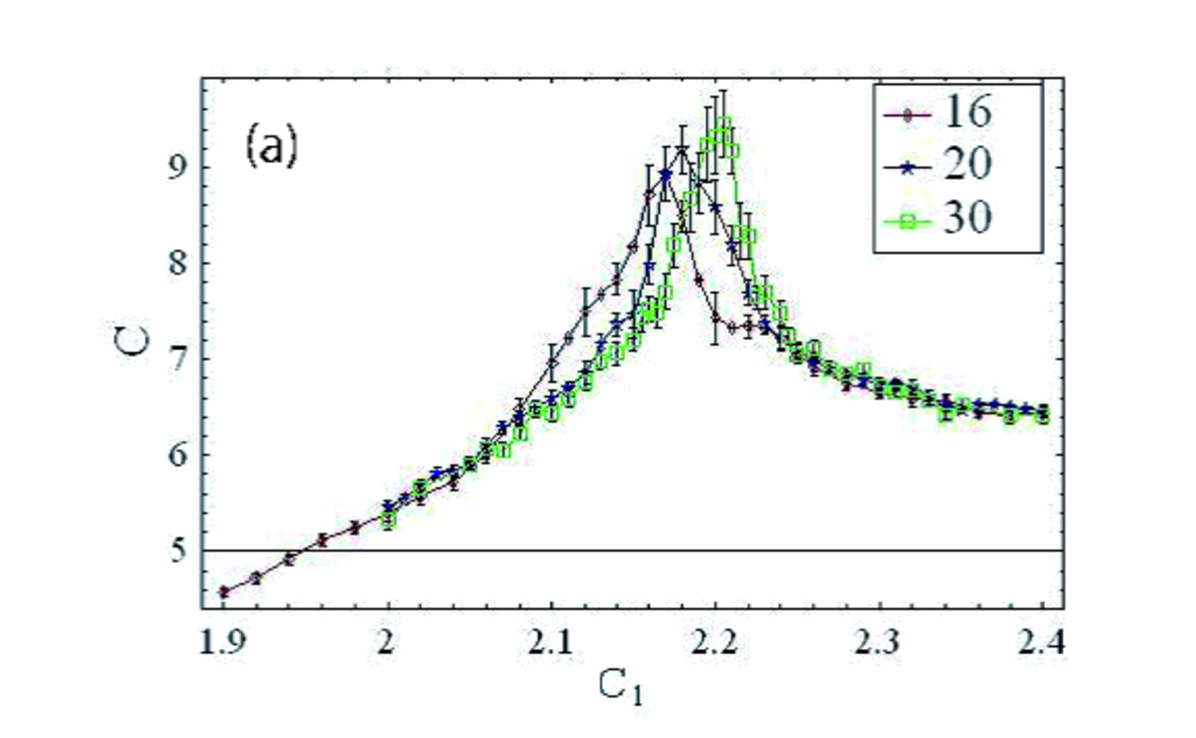}
\includegraphics[width=8cm]{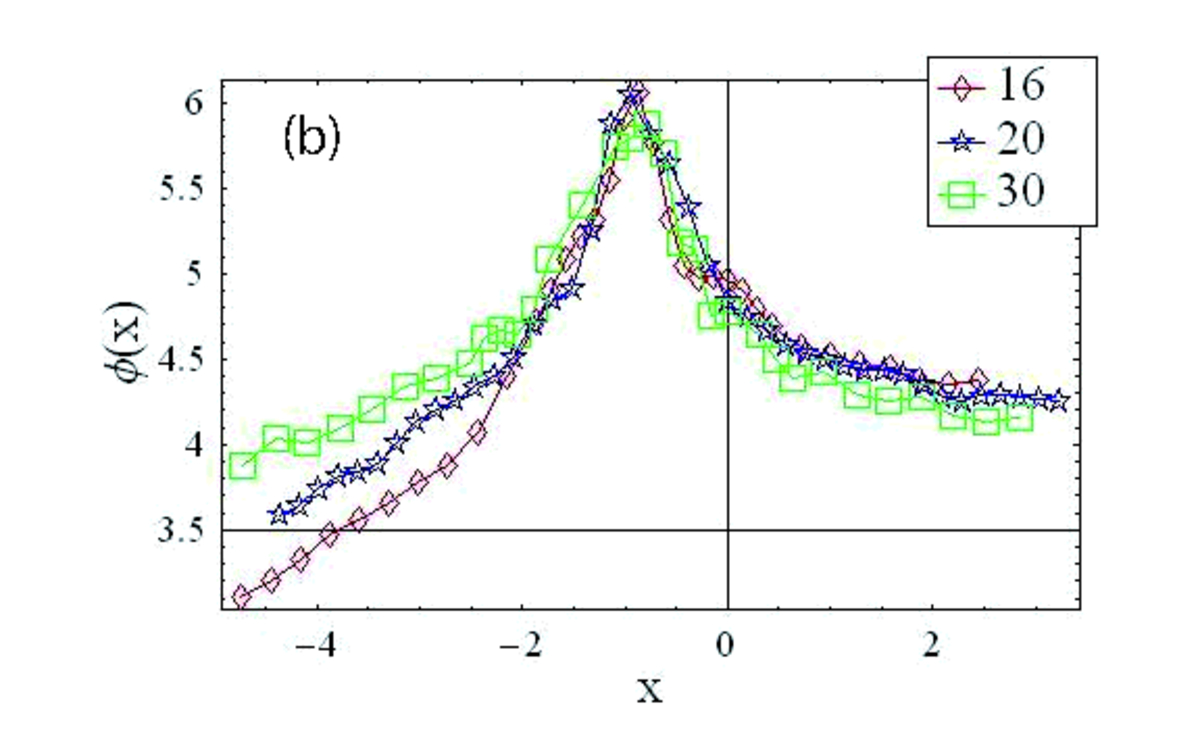}
\end{center}
\caption{(a)$C$ for $c_2=2.0,\; c_3=0$. System size is $L=16,\; 20,\; 30$.
(b)Scaling function $\phi(x)$ in finite-size scaling (\ref{fss}).
Critical exponents and critical coupling are estimated as 
$\nu=0.8,\; \sigma=0.11$ and $c_{1c}=2.23$.}
\label{ca}
\end{figure}

We first consider the case of the pure CP$^3$ model with $c_2=c_3=c_4=0$,
which corresponds to the anisotropic SU(4) AF magnet\cite{kawashima}.
We calculate the ``internal energy" $E=\langle A_L \rangle/L^3$ and
the ``specific heat" $C=\langle (A_L-E)^2\rangle/L^3$ to study phase
structure, where $L^3$ is the lattice size and we impose the periodic
boundary condition in moose of calculations.
In Fig.\ref{eb}, we show $E$ and $C$ as a function of $c_1$.
It is obvious that $E$ has a discontinuity at $c_1 \simeq 4.5$
and $C$ has a very large peak at $c_1 \simeq 4.5$, which indicates
a first-order phase transition.
In order to verify this observation, we calculated density of states $N[E]$
that is defined as 
\begin{equation}
N[E]=\int [DU][DzD\bar{z}]_{\rm CP}\; \delta (A_L-E) \; e^{A_L}.
\label{ne}
\end{equation}
Result in Fig.\ref{histoa} shows that $N[E]$ has a double-peak shape at $c_1=4.498$, whereas
it has a single peak at the other couplings.
This confirms the existence of the first-order phase transition in the 
CP$^3$ model, though corresponding phase transitions in the
CP$^1$ and CP$^2$ models are of second order.
Study of the correlation functions of the ``spin" operators given later on
shows that phase transition from ``N\'eel" to paramagnetic states takes place
at $c_1=4.498$.

CP$^{N-1}$ model in the 3D spacetime was studied by the $1/N$-expansion and 
it was suggested that there existed a second-order phase transition
from ordered to disordered phases as the coupling constant is increased.
However the present investigation by means of the MC simulations shows that
the order of the phase transition varies as a function of the parameter $N$.
Similar phenomenon was recently observed some related model, e.g. multi-Higgs U(1)
gauge model in 3D.
We also studied finite but small $c_2$ cases and found that the phase transition
is still of first-order.
However at intermediate value of $c_2$, $C$ exhibits the finite-size scaling
(see Fig.\ref{ca}), i.e., the data of $C(c_1, L)$ for system size $L$ and 
$c_2=2.0$ can be fit as follows with a scaling function $\phi(x)$\cite{fss},
\begin{equation}
C(c_1, L)=L^{\sigma/\nu}\phi(L^{1/\nu}\epsilon), \;\;\;
\epsilon=(c_1-c_{1c})/c_{1c},
\label{fss}
\end{equation}
where $c_{1c}$ is the critical coupling at infinite system size
and estimated as $c_{1c}=2.23$.
This fact means that the phase transition becomes of second-order as 
the value of $c_2$ is increased.

\begin{figure}[ht]
\begin{center}
\includegraphics[width=7.8cm]{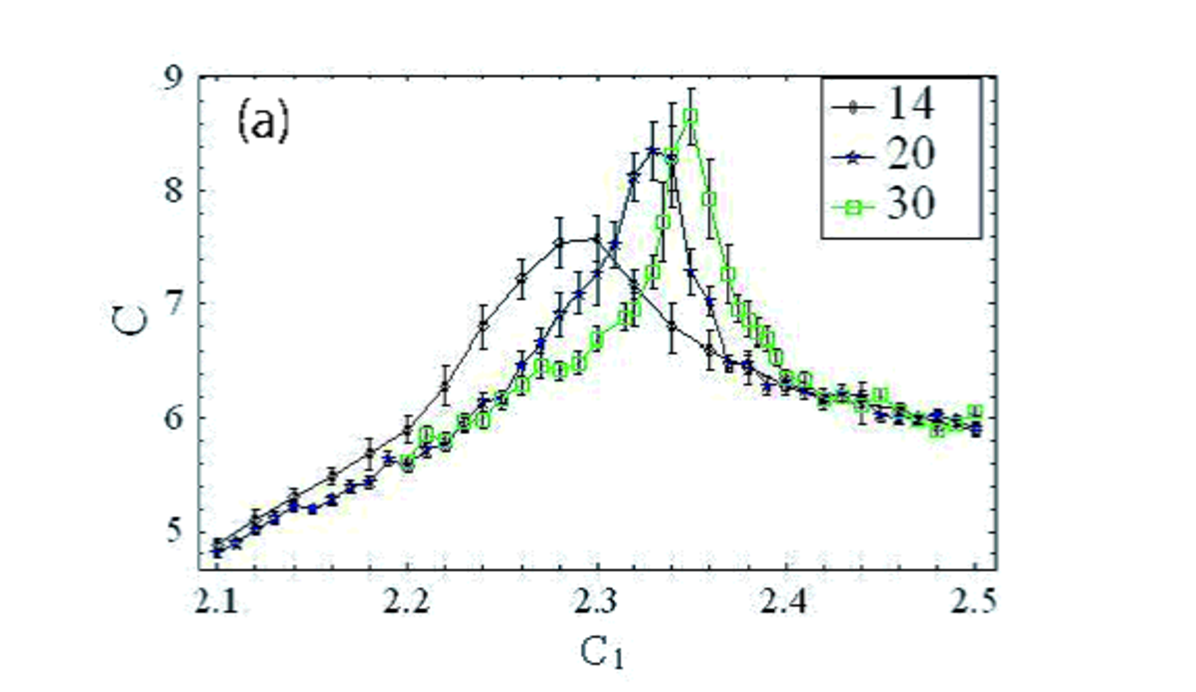}
\includegraphics[width=7cm]{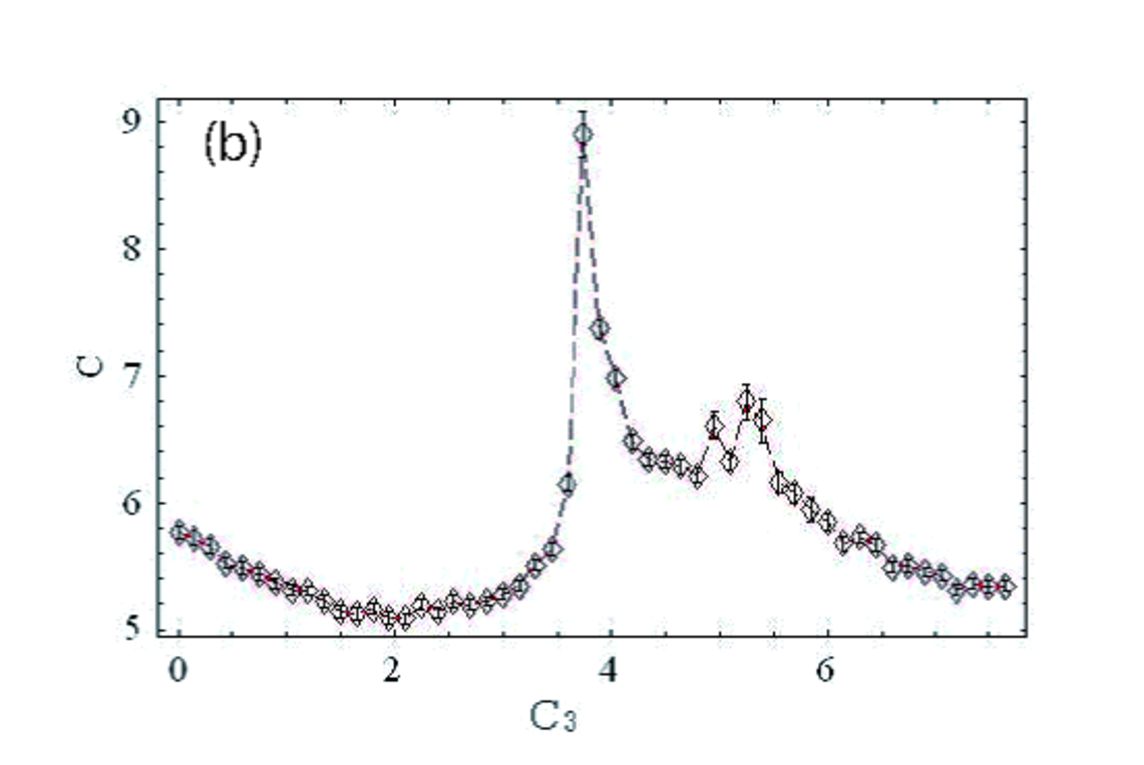}
\end{center}
\caption{(a)$C$ for $c_2=2.0,\; c_3=1.0$. System size is $L=14,\; 20,\; 30$.
Critical coupling is estimated at $c_1=2.35$.
(b)$C$ for $c_1=3.0,\; c_2=2.0$. System size is $L=18$.
There are two phase transitions at $c_3=3.7$ and $5.2$.}
\label{cb}
\end{figure}

Let us turn on the $c_3$-term and see how location of the phase transition varies.
We studied the system by varying value of $c_1$ with fixed $c_3$, 
and found clear signal of phase transitions, see, e.g. $C$ for 
$c_2=2.0, \; c_3=1.0$ in Fig.\ref{cb}.
We also investigated phase structure of the system with $c_1$ fixed and $c_3$ varied,
and found that there is another phase transition line.
See Fig.\ref{cb}.
\begin{figure}[ht]
\begin{center}
\includegraphics[width=8cm]{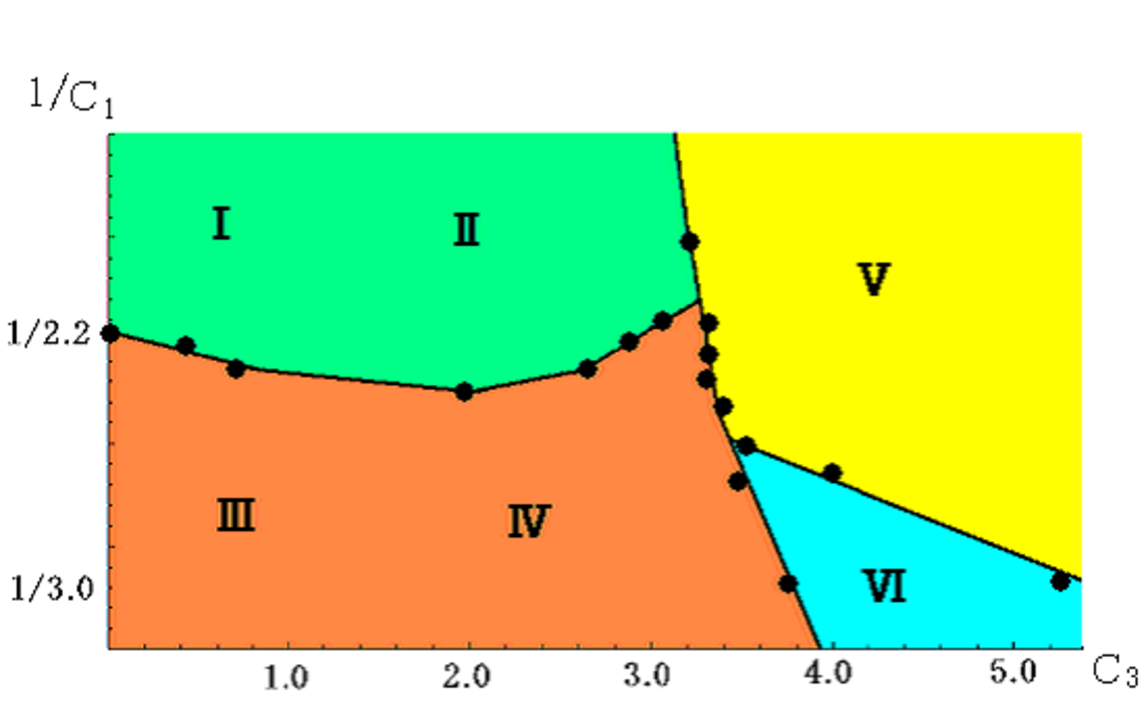}
\end{center}
\caption{Phase diagram in $(c_3-{1 \over c_1})$ plane for $c_2=2.0$.
Solid lines are phase transition lines obtained by measurement of $E$ and $C$.
Dots denote phase transition points actually observed by measurement.
``Spin" correlation functions exhibit different behavior in regions
$I\sim VI$. Similar phase structure is obtained for $c_2=0$.}
\label{phase}
\end{figure}
\begin{figure}[ht]
\begin{center}
\includegraphics[width=6cm]{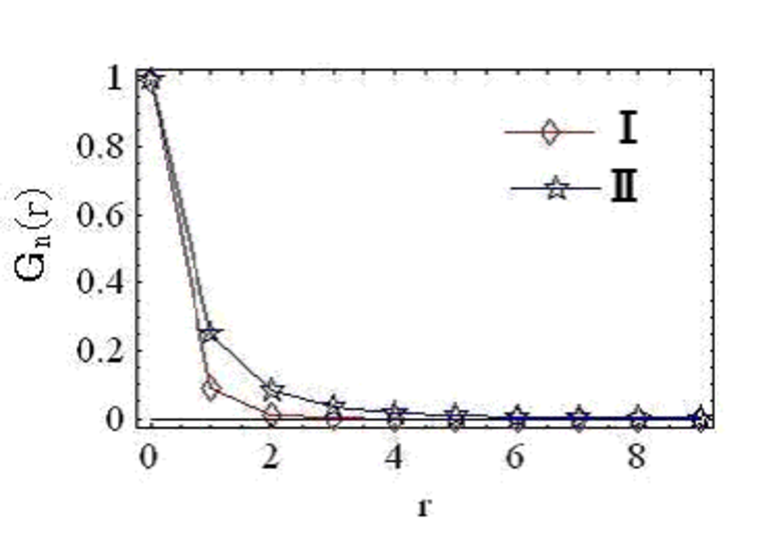}
\includegraphics[width=6cm]{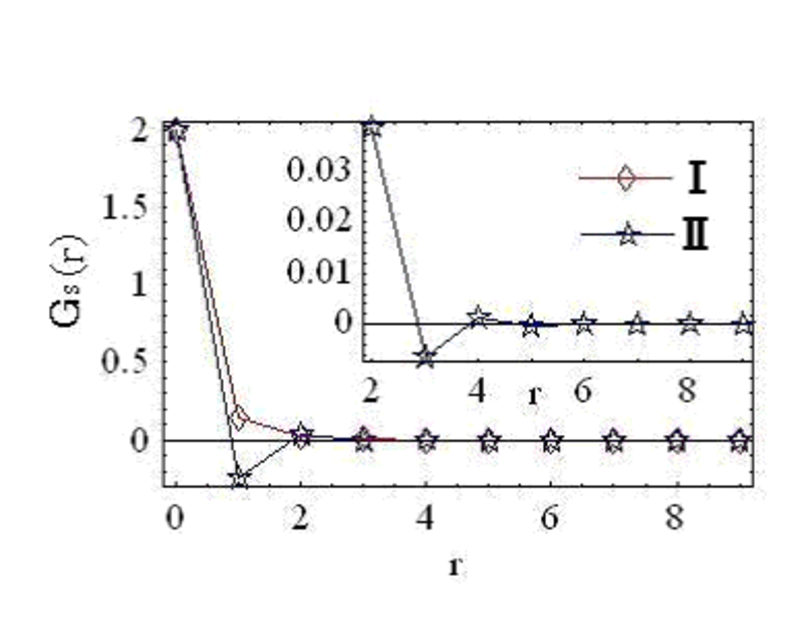}
\end{center}
\caption{Spin correlation functions in $I$ and $II$ in phase diagram 
Fig.\ref{phase}.
Both $G_n(r)$ and $G_s(r)$ have no long-range order in $I$ and $II$, however
in $II$ $G_s(r)$ exhibits a short-range spiral order.}
\label{spincora}
\end{figure}
\begin{figure}[ht]
\begin{center}
\includegraphics[width=6cm]{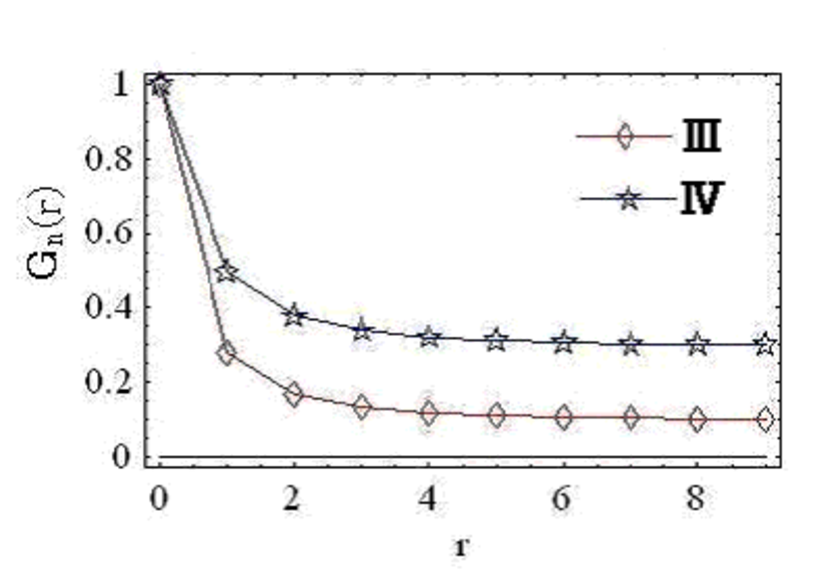}
\includegraphics[width=6cm]{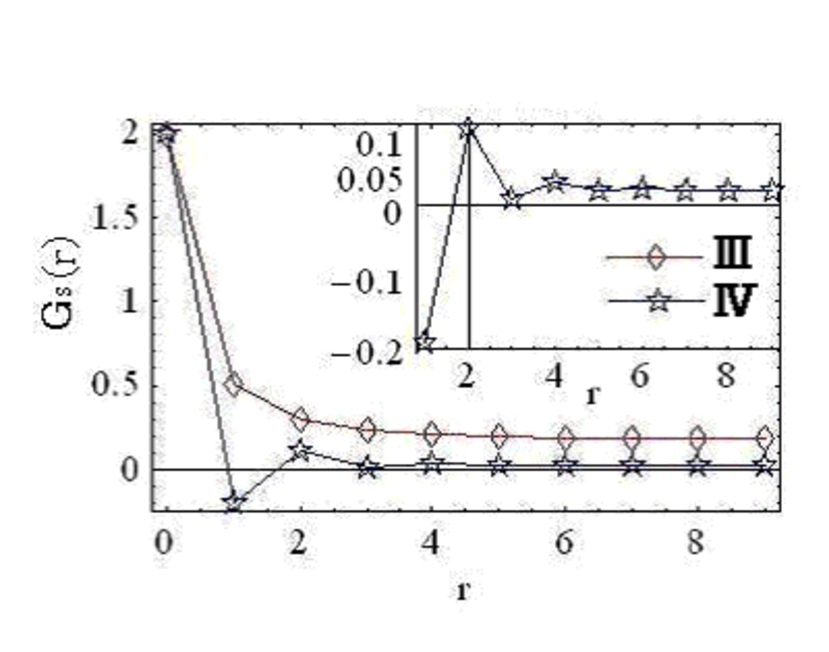}
\end{center}
\caption{Spin correlation functions in $III$ and $IV$ in phase diagram 
Fig.\ref{phase}.
$G_n(r)$ has long-range order both in $III$ and $IV$.
In $IV$, $G_s(r)$ exhibits a long-range spiral order, whereas it has 
usual long-range order in $III$.}
\label{spincorb}
\end{figure}
\begin{figure}[bh]
\begin{center}
\includegraphics[width=6cm]{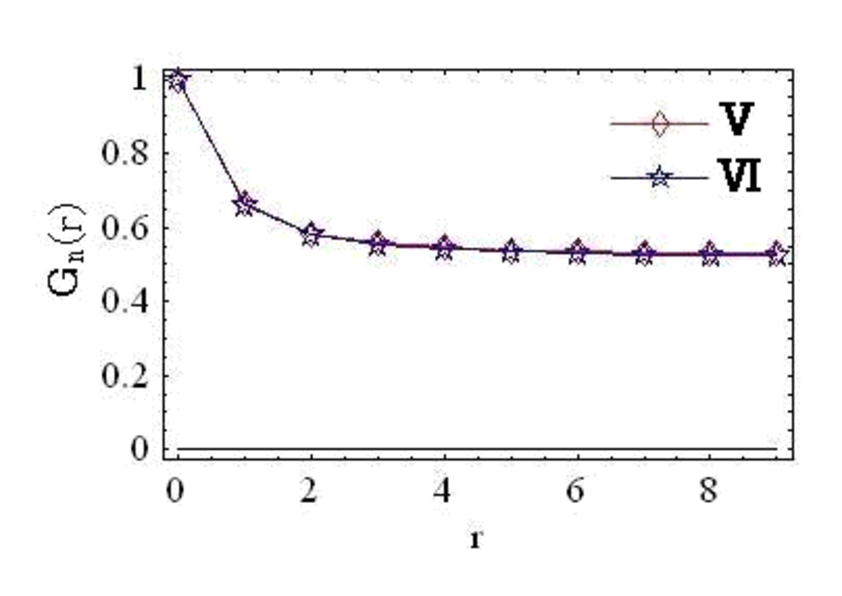}
\includegraphics[width=6cm]{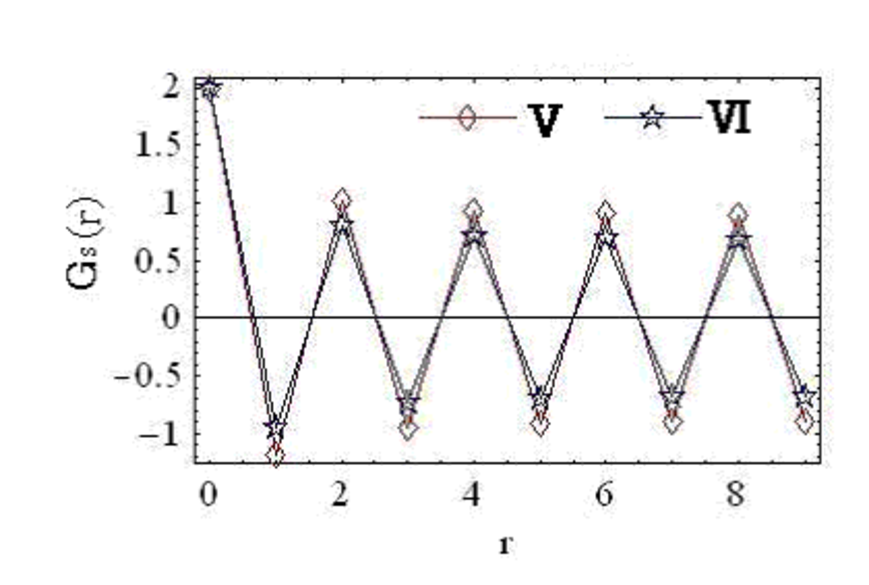}
\end{center}
\caption{Spin correlation functions in $V$ and $VI$ in phase diagram 
Fig.\ref{phase}. Both $G_n(r)$ and $G_s(r)$ exhibit the same behavior
in $V$ and $VI$.
}
\label{spincorc}
\end{figure}
\begin{figure}[ht]
\begin{center}
\includegraphics[width=8cm]{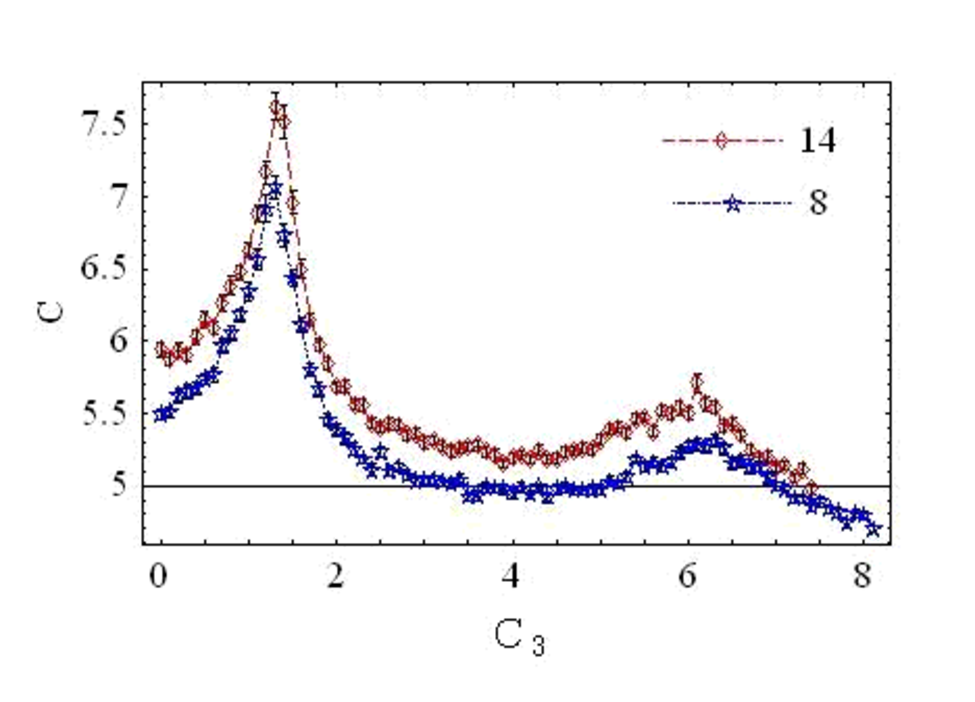}
\end{center}
\caption{Specific heat on line $c_3=c_4$ with $c_1=6,\; c_2=0$.
Result indicates two phase transitions on the line.
}
\label{cd}
\end{figure}
\begin{figure}[ht]
\begin{center}
\includegraphics[width=6.5cm]{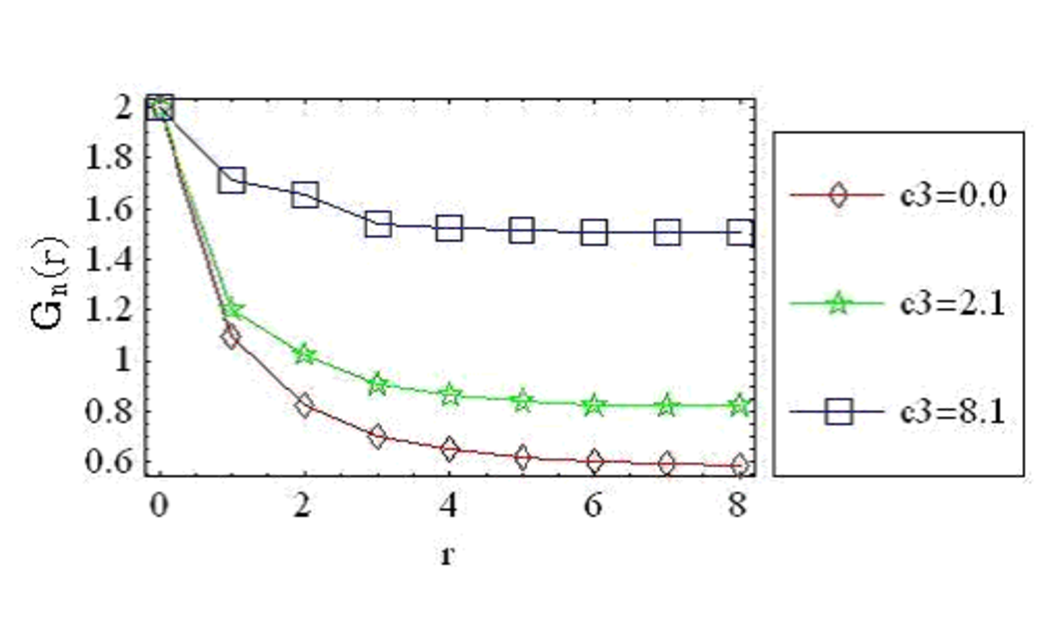}
\includegraphics[width=6.5cm]{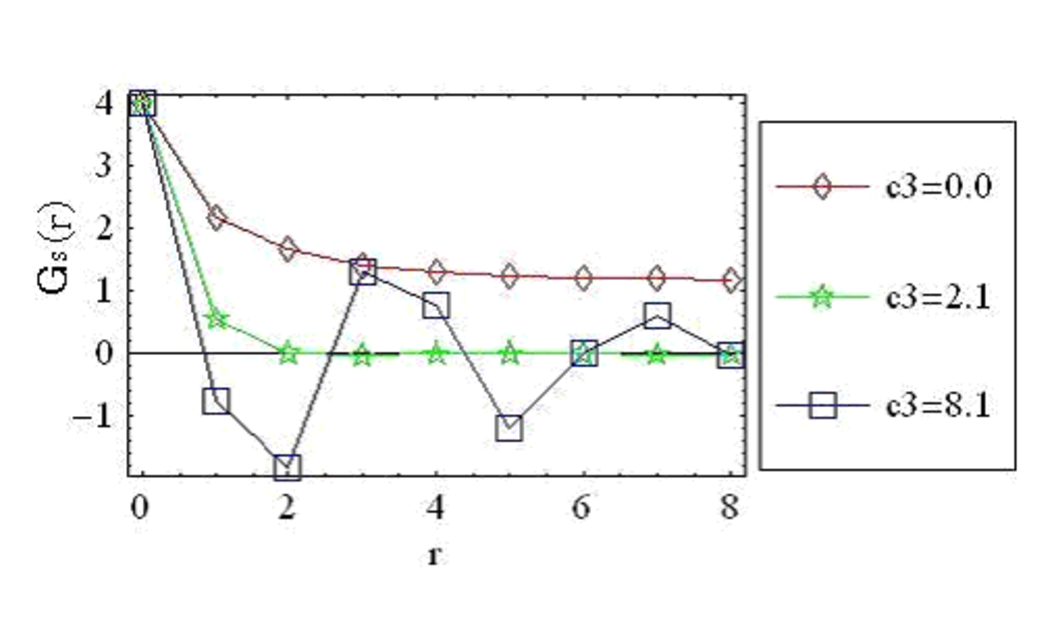}
\end{center}
\caption{Spin correlation functions for $c_1=6.0$ and $c_2=0$.
}
\label{spincord}
\end{figure}

Obtained phase diagram in the $(c_3-{1 \over c_1})$ plane is shown in Fig.\ref{phase}.
There are four phases, which are identified by the measurement of $E$ and $C$.
We also investigated the behavior of the correlation function of the ``spin" operators,
\begin{equation}
G_n(r)=\sum_a\langle\Gamma^a_{r'+r}\Gamma^a_{r'} \rangle, \;\;
G_s(r)=\sum_{ab}\langle\Gamma^{ab}_{r'+r}\Gamma^{ab}_{r'} \rangle.
\label{spincorrelation}
\end{equation}
As we expect that a phase transition to a ``spiral state" takes place
as the parameters are increased, we take the free boundary condition in the
two spatial directions.
We found that the correlators exhibit different behavior in the six ``regions" 
$I\sim VI$ shown in Fig.\ref{phase}.
It is obvious that not only simple ferromagnetic(FM) correlation,  
antiferromagnetic(AF) correlations appear in these correlators.
For example in the regions $I$ and $II$ ($III$ and $IV$), the correlation of
the nematic order $G_n(r)$ exhibits the same behavior, but  
the spin correlator $G_s(r)$ behaves differently in $I$ and $II$ ($III$ and $IV$), 
see Fig.\ref{spincora} (Fig.\ref{spincorb}).
We have observed no phase boundary between the $I$ and $II$ 
($III$ and $IV$) regions on which 
the internal energy $E$ and specific heat $C$ exhibit anomalous behavior. 
However from the result of the spin correlation function, $G_s(r)$,
we expect that the phases 
$\langle \omega_\mu \rangle\neq 0, \;\langle \xi(r)\rangle=0$ and 
$\langle \omega_\mu \rangle\neq 0, \;\langle \xi(r)\rangle\neq 0$
are realized in the regions $II$ and $IV$, respectively.
On the other hand, there are no phases in the effective field theory that
correspond to the phases $V$ and $VI$ of the FM long-range order in the
lattice model.
 
Finally we turn on the $c_4$-term in addition to the $c_3$-term.
Numerical study of the system was performed along the line $c_3=c_4$ with $c_1=6.0$
and $c_2=0$ in the phase diagram.
Observed $C$ as a function of $c_3=c_4$ is shown in Fig.\ref{cd}.
There are two phase transition, one at $c_3=c_4\simeq 0.7$ and 
the other at $c_3=c_4\simeq 6.0$.
In order to understand these phase transitions, we calculated the spin 
correlation functions in each phase, which are shown in Fig.\ref{spincord}.
From the result, it is obvious that the AF order disappears first
and then at the second transition the spiral state appears.
This state corresponds to the state with nonvanishing 
$\langle \xi(r)\rangle\neq 0$ studied in the previous sections.


\section{Conclusion}

In the present paper, we have studied the Sp(N) AF Heisenberg model
in 2D that is expected to be realized in cold atom system in an 
optical lattice.
We focus on the ground state structure of the system and derived
the effective field theory for the system, which is a kind of extension
of the CP$^{N-1}$ nonlinear $\sigma$-model.
Then we studied the phase structure and critical behavior by using the
$1/N$-expansion.
We found that the spatial anisotropy induces a phase transition from the
ordered state to paramagnetic state.
As the explicit breaking of the SU(N) symmetry increases, the ground state exhibits
a spiral order of the ad joint representation of Sp(N) group.
This state also has a nematic order of vector representation of the Sp(N)
group.

We also introduced a lattice gauge-model counterpart of the effective
field theory, and studied its phase structure by means of the MC simulations.
We found a similar phase diagram with that of the field theory,
but order of the phase transitions is different in two systems.

In Ref.\cite{xu}, finite-temperature phase diagram of Sp(N) model is studied
by using a Ginzburg-Landau theory in terms of gauge-invariant spin
fields, $\phi^{ab}=z^\dagger \Gamma^{ab}z, \; \phi^a=z^\dagger \Gamma^{ab}z$.
There quantum phase transition of Sp(4) spin system in 3D stacked square
lattice is also discussed.
Phase transition from CP$^3$ N\'eel ordered state to photon liquid state
is predicted.
In the (3+1)D counterpart of the lattice model (5.1), it is expected
that deconfined photon phase exists for sufficiently large $c_2$.
U(1) gauge theory of CP$^1$ field in (3+1)D was studied previously\cite{sawamura}, 
and it was found that phase transition from CP$^1$ N\'eel ordered state 
to photon liquid state is of second order.

It is interesting to study effects of hole doping to the Sp(N) AF
magnets and investigate how the long-range orders are broken
and if a new phase with hole-pair condensation appears.
This system is an extension of the t-J model for high-temperature
superconductivity.
This problem is under study and we hope that results will be reported
in a future publication.

\bigskip
\begin{center}
{\bf Acknowledgment} \\
\end{center}
This work was partially supported by Grant-in-Aid
for Scientific Research from Japan Society for the 
Promotion of Science under Grant No.20540264.


\newpage
\appendix
\begin{center}
\end{center}
\normalsize
\section{Derivation of the effective field theory}
\renewcommand{\theequation}{A.\arabic{equation}} 
\setcounter{equation}{0}
In Sec.2, we derived the effective field theory of the Sp(N)
AF Heisenberg model.
In the present appendix, we give some details of the derivation.

In Eq.(\ref{aaa}), $A(\tau)$ is expanded in powers of ${\bf p}$
and $\bar{\bf p}$.
Vectors ${\bf k}$ and $\bar{\bf l}$ are explicitly given as 
follows,
\begin{align}
\bf{k} = \left( \begin{array}{c}
-U(\bar{w}_{0}\dot{w}_{1})
-\displaystyle\sum_{j}J_{o,j}(\bar{z}_{j}\mathcal{J}\bar{w}_{0})
(z_{j}\mathcal{J}w_{1})U
-\displaystyle\sum_{j}J'_{o,j}(\bar{w}_{0}z_{j})(\zb_{j}\w)U \\
-U(\bar{w}_{2}\dot{w}_{1})
-\displaystyle\sum_{j}J_{o,j}(\bar{z}_{j}\mathcal{J}\bar{w}_{2})
(z_{j}\mathcal{J}w_{1})U
-\displaystyle\sum_{j}J'_{o,j}(\bar{w}_{2}z_{j})(\zb_{j}\w)U \\
-U(\bar{w}_{3}\dot{w}_{1})
-\displaystyle\sum_{j}J_{o,j}(\bar{z}_{j}\mathcal{J}\bar{w}_{3})
(z_{j}\mathcal{J}w_{1})U 
-\displaystyle\sum_{j}J'_{o,j}(\bar{w}_{3}z_{j})(\zb_{j}\w)U 
\label{vec_k}
\end{array}
\right)
\end{align}
\begin{equation}
\bar{\bf{l}} = \left( \begin{array}{c}
-\bar{U}(\bar{w}_{1}\dot{w}_{0})
-\displaystyle\sum_{j}J_{o,j}(z_{j}\mathcal{J}w_{0})
(\bar{z}_{j}\mathcal{J}\bar{w}_{1})\bar{U}
-\displaystyle\sum_{j}J'_{o,j}(\zb_{j}w_{0})(\wb z_{j})\Ub \\
-\bar{U}(\bar{w}_{1}\dot{w}_{2})
-\displaystyle\sum_{j}J_{o,j}(z_{j}\mathcal{J}w_{2})
(\bar{z}_{j}\mathcal{J}\bar{w}_{1})\bar{U} 
-\displaystyle\sum_{j}J'_{o,j}(\zb_{j}w_{2})(\wb z_{j})\Ub \\
-\bar{U}(\bar{w}_{1}\dot{w}_{3})
-\displaystyle\sum_{j}J_{o,j}(z_{j}\mathcal{J}w_{3})
(\bar{z}_{j}\mathcal{J}\bar{w}_{1})\bar{U} 
-\displaystyle\sum_{j}J'_{o,j}(\zb_{j}w_{3})(\wb z_{j})\Ub 
\label{vec_l}
\end{array}
\right)
\end{equation}
After integrating out $\mathbf{p}$ and $\bar{\bf p}$,
we obtain Eq.(\ref{ac}).
For $J'_{o,j} \ll J_{oj}$, the inverse of the matrix ${\bf M}$
can be approximated as 
\begin{align}
{\bf M}^{-1} 
&= \left( \begin{array}{ccc}
-(4J_{o,j}+4J'_{o,j})^{-1} & 0                & 0 \\
0                          & -(4J_{o,j})^{-1} & 0 \\
0                          & 0                & -(4J_{o,j})^{-1}
\end{array}
\right) \notag\\
&\approx
\left( \begin{array}{ccc}
-\frac{1}{4J_{o,j}}+\frac{J'_{o,j}}{4J_{o,j}^{2}} & 0                   & 0 \\
0                                                 & -\frac{1}{4J_{o,j}} & 0 \\
0                                                 & 0                   & -\frac{1}{4J_{o,j}}
\end{array}
\right) \label{Inv_M}
\end{align}
and
\begin{align}
A_{z}(\tau)
&=\sum_{\mathrm{odd}}\bigg[
-\sum_{k}{'}\frac{1}{2J_{k}}(\bar{w}_{k}\dot{w}_{1})(\bar{w}_{1}\dot{w}_{k})    \notag \\
&-J_{o,i}\sum_{k}{'}\frac{1}{2J_{k}}\sum_{i} 
\{(\bar{z}_{i}\mathcal{J}\bar{w}_{k})(z_{i}\mathcal{J}w_{1})
(\bar{w}_{1}\dot{w}_{k}) 
+(z_{i}\mathcal{J}w_{k})(\bar{z}_{i}\mathcal{J}\bar{w}_{1})
(\bar{w}_{k}\dot{w}_{1})\} \notag \\
&-J^{2}_{o,i}\sum_{k}{'}\frac{1}{2J_{k}}\sum_{i,j}
(\bar{z}_{i}\mathcal{J}\bar{w}_{k})(z_{i}\mathcal{J}w_{1})
(z_{j}\mathcal{J}w_{k})(\bar{z}_{j}\mathcal{J}\bar{w}_{1}) \notag \\
&-J'_{o,i}\sum_{k}{'}\frac{1}{2J_{k}}\sum_{i}
\{(\bar{w}_{k}z_{i})(\bar{z}_{i}w_{1})(\bar{w}_{1}\dot{w}_{k})
+(\bar{z}_{i}\bar{w}_{k})(\bar{w}_{1}z_{i})(\bar{w}_{k}\dot{w}_{1})\} \notag \\
&-J_{o,i}J'_{o,i}\sum_{k}{'}\frac{1}{2J_{k}}\sum_{i,j}
\{(\bar{z}_{i}\mathcal{J}\bar{w}_{k})(z_{i}\mathcal{J}w_{1})
(\bar{z}_{j}w_{k})(\bar{w}_{1}z_{j}) \notag \\
&+(z_{i}\mathcal{J}w_{k})(\bar{z}_{i}\mathcal{J}\bar{w}_{1}) 
(\bar{w}_{k}z_{j})(\bar{z}_{j}w_{1})\} \notag \\
&-(J'_{o,i})^2\sum_{k}{'}\frac{1}{2J_{k}}\sum_{i,j}
(\bar{w}_{k}z_{i})(\bar{z}_{i}w_{1})
(\bar{z}_{j}w_{k})(\bar{w}_{1}z_{j})  \notag \\
&-J_{o,i}\sum_{i}(\bar{z}_{i}\mathcal{J}\bar{w}_{1})(z_{i}\mathcal{J}w_{1})
-J'_{o,i}\sum_{i}(\wb z_{i})(\zb_{i}\w) \bigg],
\label{eqbd}
\end{align}
where $\sum_{k}{'}$ denotes the sum over $k=0,2,3$, and 
$J_{0}\approx 4J_{o,i} + 4J'_{o,i}$A$J_{2} = J_{3} = 4J_{o,i}$.
By using identity such as
\begin{eqnarray}
&& (\zb_{i}z_{j})=(\zb_{i}z)(\zb z_{j}) + a^2 (\Db_{i}\zb D_{j}z), \nonumber  \\
&& D_{\mu} = \partial_{\mu} + iA_{\mu}, \;\;
 A_{\mu} = i\zb \partial_{\mu}z, 
\label{eqbe}
\end{eqnarray}
where $a$ is the lattice spacing,
and the completeness of $\{w_k\}$,
\begin{equation}
\sum_{k=0}^{3}w_{k\alpha}\bar{w}_{k\beta}=\delta_{\alpha\beta}, 
\end{equation}
we obtain
\begin{align}
A_{z}(\tau)=&\sum_{\mathrm{odd}} \bigg[
\frac{1}{2Jd}\bar{D}_{\tau}\zb D_{\tau}z
-\frac{Ja^{2}}{2d}\sum_{i,j}\bar{D}_{i}\zb D_{j}z \notag\\
&-\frac{a}{2d}\sum_{j}
(\bar{D}_{\tau}\zb D_{j}z-\bar{D}_{j}\zb D_{\tau}z)
+Ja^{2}\sum_{j}\bar{D}_{j}\zb D_{j}z-2Jd \notag \\
&-\frac{J'}{2J^{2}d}(\zb\Jm \bar{D}_{\tau}\zb)(z\Jm D_{\tau}z)
+\frac{J'a^{2}}{2d}\sum_{i,j}(\zb\Jm \bar{D}_{i}\zb)(z\Jm D_{j}z) \notag\\
&-J'a^{2}\sum_{j}(\zb\Jm \bar{D}_{j}\zb)(z\Jm D_{j}z) \notag\\
&-\frac{J'}{2Jd}\sum_{j}
\{(\zb\Jm \bar{D}_{j}\zb)(z\Jm D_{\tau}z) - (\zb\Jm \bar{D}_{\tau}\zb)(z\Jm D_{j}z)\} \notag\\
&+\frac{J'}{2Jd}\sum_{j}
\{(\zb\Jm \bar{D}_{j}\zb)(z_{j}\Jm D_{\tau}z)
-(z\Jm D_{j}z)(\zb_{j}\Jm \Db_{\tau}\zb)\} \label{eqbg}
\bigg].
\end{align}
From Fig.\ref{ab}, by substituting 
\begin{align}
\begin{cases}
D_{1}=(-)^yD_{x}+D_{y} \\
D_{2}=2(-)^yD_{x} \\
D_{3}=(-)^yD_{x}-D_{y}, \label{eqbh}
\end{cases}
\end{align}
the effective action $S_{\rm E}$ defined by
\begin{equation}
S_{\rm E} = \int_{0}^{\beta}d\tau A_{z}(\tau) 
\label{se}
\end{equation}
is obtained as 
\begin{align}
S_{\rm E}&=\int_{0}^{\beta}d\tau
\bigg[\frac{1}{4J}\Db_{\tau}\zb D_{\tau}z 
+2Ja^{2}\sum_{\mu =x,y}\Db_{\mu}\zb D_{\mu}z 
-a(-)^y(\Db_{\tau}\zb D_{x}z - \Db_{x}\zb D_{\tau}z) \notag\\
&-\frac{J'}{4J^{2}}(\zb\Jm \bar{D}_{\tau}\zb)(z\Jm D_{\tau}z)
-2J'a^{2}\sum_{\mu=x,y}(\zb\Jm \bar{D}_{\mu}\zb)(z\Jm D_{\mu}z) \notag\\
&-\frac{J'}{J}(-)^y
\{(\zb\Jm \bar{D}_{x}\zb)(z\Jm D_{\tau}z) - 
(\zb\Jm \bar{D}_{\tau}\zb)(z\Jm D_{x}z)\} \notag\\
&+\frac{J'}{4J}\sum_{j}
\{(\zb\Jm \bar{D}_{j}\zb)(z_{j}\Jm D_{\tau}z)
-(z\Jm D_{j}z)(\zb_{j}\Jm \Db_{\tau}\zb)\} 
\bigg]. \label{seb}
\end{align}
By the rescaling (\ref{rescaling}), 
\begin{align}
S &= \frac{\sqrt{2\lambda}}{2a(1+\lambda)}\int d^{3}r
\bigg[
\sum_{\mu=\tau,x,y}\Db_{\mu}\zb D_{\mu}z 
-\sqrt{1+\lambda}a\partial_y(\Db_{\tau}\zb D_{x}z -\Db_{x}\zb D_{\tau}z ) \notag\\
&-\frac{J'_{0}}{J_{0}}\sum_{\mu=\tau,x,y}
(\zb \Jm \Db_{\mu}\zb)(z\Jm D_{\mu}z)
-\frac{J'_{0}}{J_{0}}\sqrt{1+\lambda}a\partial_y(\Db_{x}\zb D_{\tau}z 
-\Db_{\tau}\zb D_{x}z ) \notag\\
&+\sum_{j}\frac{J'_{j}}{4J_{0}}\sqrt{1+\lambda}
\{(\zb \Jm \Db_{j}\zb)(z_{j}\Jm D_{\tau}z)
-(z\Jm D_{j}z)(\zb_{j}\Jm \Db_{\tau}\zb)\} \bigg]. \label{sec}
\end{align}
The last term in Eq.(\ref{sec}) is rewritten as 
\begin{align}
&\sum_{j}\frac{J'_{j}}{4J_{0}}\sqrt{1+\lambda}
\{(\zb \Jm \Db_{j}\zb)(z_{j}\Jm D_{\tau}z)
-(z\Jm D_{j}z)(\zb_{j}\Jm \Db_{\tau}\zb)\} \notag\\
&\rightarrow
\sum_{j}\frac{J'_{j}}{4J_{0}}\sqrt{1+\lambda}
\{(\zb \Jm \Db_{j}\zb)(z\Jm D_{\tau}z)
-(z\Jm D_{j}z)(\zb\Jm \Db_{\tau}\zb)\} \notag\\
&=\frac{J'_{0}}{J_{0}}\sqrt{1+\lambda}(\Db_{x}\zb D_{\tau}z -\Db_{\tau}\zb D_{x}z ), \label{eqcd}
\end{align}
then,
\begin{align}
S_{\rm E} = \frac{\sqrt{2\lambda}}{2a(1+\lambda)}\int d^{3}r
\bigg[
\sum_{\mu=\tau,x,y}\Db_{\mu}\zb D_{\mu}z 
-\frac{J'_{0}}{J_{0}}
\sum_{\mu=\tau,x,y}(\zb \Jm \Db_{\mu}\zb)(z\Jm D_{\mu}z) \bigg]
+S_{B}, \label{SE4}
\end{align}
where 
$S_{B}$ is the Berry phase,
\begin{equation}
S_{B} = -\frac{1}{2}\sqrt{\frac{2\lambda}{1+\lambda}}
\Big(1-{J'_0 \over J_0}\Big)
\int d^{3}rD_y
(\Db_{\tau}\zb D_{x}z -\Db_{x}\zb D_{\tau}z ). \label{bpa}
\end{equation}






\begin{thebibliography}{99}
\bibitem{fnx}
For SU(N) (N$\le 4$) Heisenberg model with isotropic nearest-neigbhor coupling, 
it is known that the ground state has the AF
long-range order\cite{kawashima}.
This means that effective coupling of these spin models are smaller than 
the critical coupling $g_c$.
\bibitem{fna}
However
as the broken symmetry of Sp(N) in the phases of $\omega=0$ 
and $\omega \neq 0$ is defferent, a first-order
phase transition is possible.
\bibitem{fnb}Here we should notice that the lattice gauge model
$A_L$ is a compact U(1) gauge model as the original Sp(N) spin model
with the Schwinger-boson representation has the compact U(1)
gauge symmetry.
\end{thebibliography}
\end{document}